\documentclass[aps,physrev,twocolumn,showpacs,superscriptaddress,amsmath,amssymb,amsfonts,floatfix]{revtex4-2}

\usepackage[utf8]{inputenc}
\usepackage[T1]{fontenc}        
\usepackage{lmodern}            
\usepackage{graphicx}
\usepackage{multirow}
\usepackage{epsfig}
\usepackage{url}
\usepackage{color}
\usepackage[normalem]{ulem} 
\usepackage[colorlinks,breaklinks,bookmarks=true,citecolor=blue,linkcolor=blue,urlcolor=blue]{hyperref}
\usepackage{color}
\usepackage{tabularx}
\usepackage{sidecap}
\usepackage{float}
\usepackage{subcaption}

\newcommand{\orcidlink}[1]{}

\usepackage{amsmath}            
\usepackage{amstext}            
\usepackage{amssymb}            
\usepackage{mathtools}          
\usepackage{scalerel,stackengine} 
\usepackage{ulem}
\usepackage{xcolor}

\begin{document}

\title{Surfaces and interfaces of infinite-layer nickelates studied by dynamical mean-field theory}

\author{Leonard M. Verhoff\orcidlink{0009-0004-3358-1312}}
\affiliation{Institute of Solid State Physics, TU Wien, 1040 Vienna, Austria}

\author{Liang Si\orcidlink{0000-0003-4709-6882}}
\affiliation{School of Physics, Northwest University, Xi'an 710127, China}
\affiliation{Institute of Solid State Physics, TU Wien, 1040 Vienna, Austria}

\author{Karsten Held\orcidlink{0000-0001-5984-8549}}
\email[]{Corresponding authors: leonard.verhoff@tuwien.ac.at; held@ifp.tuwien.ac.at}
\affiliation{Institute of Solid State Physics, TU Wien, 1040 Vienna, Austria}

\begin{abstract}
Infinite-layer nickelate superconductors are typically synthesized as thin films and thus include, besides the more bulk-like inner layers, distinct surface and interface layers in contact with the vacuum and substrate, respectively. Here, we employ density-functional theory and  dynamical mean-field theory to investigate how electronic correlations influence these surface and interface regions.
Our results show that electronic correlations can significantly modify the electronic structure, even driving surface layers into a Mott-insulating state with a 3$d^8$ electronic configuration. Moreover, surface termination effects induce a polar field that can shift the $\Gamma$ and $A$ pocket above the Fermi level, even for the undoped
parent compound NdNiO$_2$.
Finally, for an $n$-type interface, often synthesized experimentally, we find the Ti 3$d$ orbitals to become electron doped.
\end{abstract}

\date{\today}

\maketitle

\section{Introduction}
The discovery of superconductivity in infinite-layer (IL) nickelate films
\cite{li2019superconductivity} has 
inspired extensive experimental~\cite{zeng2020,Wang2022,Lee2023,pan2022superconductivity,chow2025bulk} investigations and theoretical efforts~\cite{Nomura2019,Sakakibara2020,Lechermann2019,Si2020,Karp2020,Petocchi2020}, sometimes hailed as 
a new ``nickel age'' of superconductivity \cite{Norman2020,Si2022a,Chow2024}.
Despite these advances, superconductivity has yet to be realized in bulk IL nickelates, primarily due to
challenges in synthesis and chemical reduction from $R$NiO$_3$ to $R$NiO$_2$ \cite{li2020absence,Puphal2023,gainza2023evidence,hu2024atomic}.
Most theoretical studies to date are based on first principle calculations for bulk. In these studies, often the in-plane lattice constants are fixed to that of the substrate such as SrTiO$_3$ (3.905\,\AA) or (LaAlO$_3$)$_{0.3}$(Sr$_2$TaAlO$_6$)$_{0.7}$ (LSAT: 3.868\,\AA) while the out-of-plane lattice constant is relaxed within density functional theory (DFT) structural optimizations. This procedure is  detailed in Ref.~\cite{DiCataldo2023} for (Pr,Sr)NiO$_2$,  including effects of external pressure. This approach effectively models a film system of ${\cal O}(20) - {\cal O}(1000)$ atomic layers, wherein the much thicker substrate remains largely unaffected by the comparatively thin IL film.
Since the IL nickelate film is grown epitaxially, it {\it epitaxially} adopts the in-plane lattice parameters of the selected substrate. Hence, the above theoretical approximation is suitable for describing the bulk-like inner layers, but inherently neglects the deviations expected at the surface (facing vacuum) and the interface (near the substrate).
These surface and interface effects are always present but become particularly relevant in ultrathin films or heterostructures consisting of only a few nickelate layers.

  \begin{figure} [tb]
        \includegraphics[width=1\linewidth]{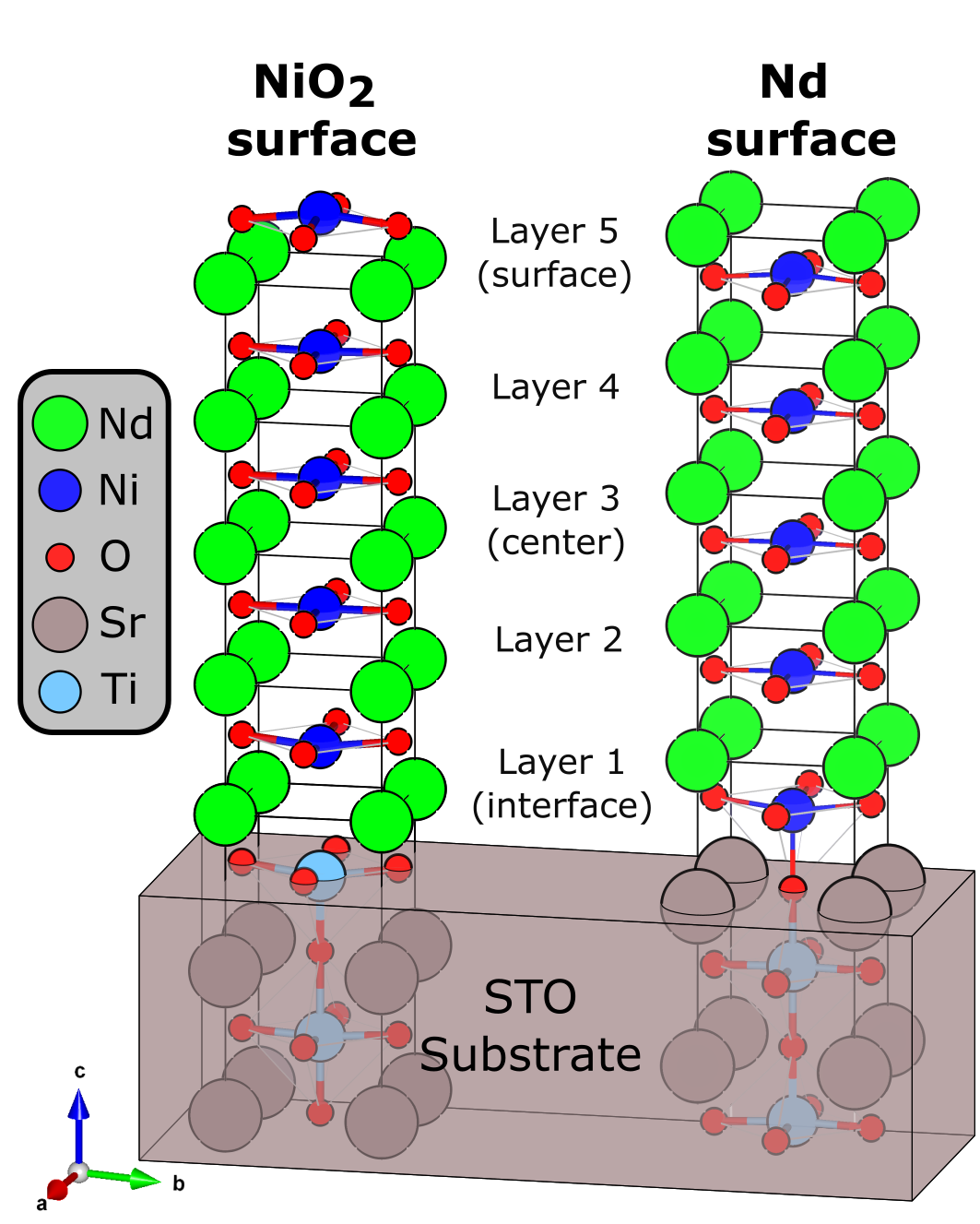}
        \caption{\label{fig:structure}DFT relaxed configurations of a NdNiO$_2$ (001) thin film, consisting of 5 NdNiO$_2$ layers on top of 2 SrTiO$_3$ (STO) layers. A stoichiometric ratio of Nd, Ni, and O gives either a NiO$_2$-terminated surface (left) or Nd-terminated surface (right).}
    \end{figure}

Interface driven polar fields have played a pivotal role in the emergence of novel two-dimensional quantum states such as an emergence of interface two-dimensional electron gas (2DEG) \cite{ohtomo2004high,Mannhart2010} and two-dimensional superconducting states \cite{reyren2007superconducting}.
Hence, interface effects in IL nickelates are expected to drive a polar field emerging from the stacking of oppositely {\it charged} $R^{3+}$ and (NiO$_2$)$^{3-}$ layers in the crystallographic $c$ direction.
For IL nickelates, polar field 
\cite{Bernardini2020,bernardini2022,hePolarityinducedElectronicAtomic2020,Geisler2021,Oritz2021} and cation intermixing \cite{Goodg2023}
have been studied  using the DFT(+$U$). 
Accurately modeling such systems requires large supercells---for example, the one shown in Fig.~\ref{fig:structure}, which contains five IL nickelate layers and two layers SrTiO$_3$ as substrate.

Including electronic correlations beyond DFT or DFT+$U$ for IL nickelate surfaces and interfaces---while feasible with advanced methods such as DMFT \cite{Georges1996,Kotliar2006,held2007electronic}---remains in its infancy.
Notably, Ref.~\cite{Oritz2021} investigated  a (LaNiO$_2$)$_8$|(LaGaO$_3$)$_4$ superlattice containing  eight NiO$_2$ layers. The study found that the six inner layers predominantly exhibit Ni 3$d_{x^2-y^2}$ character, while the interfacial Ni 3$d_{z^2}$ orbitals host additional holes.
Similar physics has also been found on the DFT level and experimentally  by the same group \cite{Ortiz2025} for the NdNiO$_2$|SrTiO$_3$ interface  that is included in our slab Fig.~\ref{fig:structure}.
These holes, together with the Ni 3$d_{x^2-y^2}$ electrons, form a high-spin state  due to Hund's exchange coupling.
Further, in Ref.~\cite{Hausoel2025}, finite-layer nickelates of the form Nd$_{n+1}$Ni$_n$O$_{2n+2}$ were investigated using DMFT and dynamical vertex approximation (D$\Gamma$A)~\cite{Toschi2007,RMPVertex}. The DMFT-caluclated electronic structures, and the critical superconducting temperature $T_\text{C}$ computed by  D$\Gamma$A demonstrate that these materials can be interpreted as almost independent 
$n$-layer-thick NdNiO$_2$ slabs, that are separated by Ruddlesden-Popper-type stacking faults.
    
Experimentally, inhomogeneous distributions of oxygen and (doped) cations, as well as structural distortions near the interface, have been studied using scanning tunneling electron microscopy (STEM),  electron energy loss spectroscopy (EELS) and x-ray diffraction (XRD) \cite{Oritz2021,Yang2023,Goodg2023,yan2024superconductivity,qi2024impact,Sahib2025}.
These investigations  generally conclude that the interface is rather abrupt.
Based on the absence of a 2DEG at the nickelate-substrate interface and detailed comparison between experiment and theory, it has been proposed that the interface is better described by a mixed intermediate NdTi$_{0.5}$Ni$_{0.5}$O$_3$ layer~\cite{Goodg2023}.
This interpretation aligns with residual oxygen signatures and Ni–Ti intermixing reported in Ref.~\cite{Yang2023}. Moreover, Ref.~\cite{Yang2023} demonstrated that the degree of interface reconstruction is thickness-dependent, likely driven by the buildup of the polar field with increasing film thickness. Additionally, Ref.~\cite{Raji2024} compared the commonly observed $n$-type interfaces with the less typical $p$-type interface found in Pr-based nickelates, concluding that interface effects are generally confined to a penetration depth of 2-3 unit cells.

Recent angular-resolved photoemission spectroscopy (ARPES) experiments~\cite{Sun2025,Ding2024}, which probe the Fermi surface of nickelate films with a high surface sensitivity, underscore the urgent need for improved modeling of both surface and interface effects.

Furthermore, recently ultrathin superconducting IL nickelate films \cite{chow2025bulk,yan2024superconductivity} have been synthesized, in some cases comprising  only two layers. This means all layers are surface or interface layers,  in sharp contrast to the earlier experiments in which the films were typically ${\cal O}(30)$  layers thick. Both ARPES and ultrathin layers call for a better understanding of potential differences between bulk-like inner layers, the surface and the interface layers.

In the present paper, we investigate two different nickelate films, using a combination of DFT and DMFT. Both structures, displayed in Fig.~\ref{fig:structure}, contain an interface, an inner bulk, and a surface part.
Our calculations reveal the emergence of a polar field that causes electronic reconstruction and cation displacements. The later is cearly visible in the true-to-scale  Fig.~\ref{fig:structure}. 

While the occupation of Ni 3$d_{x^2-y^2}$ orbitals changes continuously in DFT as a consequence of the generated polar field, DMFT reveals a more abrupt modulation at the boundary layers: for the NiO$_2$ surface layer  in the left part of Fig.~\ref{fig:structure} and for the  NiO$_2$ interface layer to SrTiO$_3$ in the right part of Fig.~\ref{fig:structure},
there is a full second hole doped into the Ni 3$d_{xz/yz}$ and the Ni 3$d_{z^2}$ orbital, respectively.
This, together with one hole in the   3$d_{x^2-y^2}$ orbital, leads to the formation of a local spin-1 moment, due to Hund's exchange. As an even more dramatic result of this change from a 3$d^9$ to  3$d^8$ electronic configuration, the surface layer for the geometry of the left part of Fig.~\ref{fig:structure} undergoes a Mott transition and becomes insulating. Similarly, but still a bit different, for the slab of the right part of Fig.~\ref{fig:structure} the interface layer toward the SrTiO$_2$ substrate is an orbital selective Mott insulator, with only the  Ni 3$d_{z^2}$ orbital but not the Ni 3$d_{x^2-y^2}$ orbital being insulating. This is in sharp contrast to the metallic behavior observed in the inner bulk-like layers with 3$d^9$ electronic configuration.

The energetically favored and experimentally more prevalent $n$-type interface corresponds to the NiO$_2$-terminated surface in our study (i.e., the left part of Fig.~\ref{fig:structure}). In this surface region, the $\Gamma$ pocket, although present in bulk-like calculations for the parent compound NdNiO$_2$ \cite{Kitatani2020}, is shifted above the Fermi level and might become invisible in ARPES measurements.

The outline of this paper is as follows: In Section~\ref{Sec:Comp}, we detail the computational methods employed in this study.
Section~\ref{Sec:DFT} presents the DFT band structure for both surface terminations.
In Section~\ref{Sec:DMFTNiO2} and Section~\ref{Sec:DMFTNd}, we show the DMFT spectra and orbital occupations for the NiO$_2$- and Nd-termination, respectively. Finally, Section~\ref{Sec:conclusion} summarizes the main finds followed by an outlook.
Additional results are provided in the Supplemental Material (SM) \cite{SM}:
Section~S1 provides details of the Wannier projection. 
Section~S2 presents DMFT self-energies, DFT and DMFT total spectral functions, Fermi surfaces, and also the layer-resolved interstitial $s$ (see also Ref.~\cite{Gu2020}) and Ti 3$d$ bands for the two slab geometries of Fig.~\ref{fig:structure}.
Section~S3 includes further studies on (A) a larger slab, (B) a system without the SrTiO$_3$ substrate, (C) different temperatures and finally calculations with a simplified tight-binding Hamiltonian in (D).

\section{Computational Details}
\label{Sec:Comp}

We calculate the electronic structure of NdNiO$_2$ films which are epitaxially grown on SrTiO$_3$, as shown in Fig.~\ref{fig:structure}. These slab models consist of a $1 \times 1 \times 5$ NdNiO$_2$ supercell. Spatial symmetry along the $z$-direction is explicitly broken by adding two SrTiO$_3$ layers on the bottom and a vacuum of at least 20\,{\AA} above the NdNiO$_2$ films. Here, the SrTiO$_3$ layers play the role of substrate in experiments, while the vacuum region separates the film from its periodic images in DFT calculations along the $z$ direction. This construction allows for the two distinct slab models shown in Fig.~\ref{fig:structure}, characterized by either:
\begin{itemize}
    \item NiO$_2$-terminated surface and Nd-terminated ($n$-type) interface (left) or
    \item Nd-terminated surface and NiO$_2$-terminated ($p$-type) interface (right).
\end{itemize}

Both supercells are stoichiometric and electrically neutral, each comprising two formula units of SrTiO$_3$ and five of NdNiO$_2$. The only difference between the two models is the type of their surface and interface terminations.
The structures are relaxed using DFT within the projector augmented-wave formalism \cite{PhysRevB.59.1758,Kresse1994} as implemented in \textsc{vasp} \cite{PhysRevB.47.558,Kresse1996}. Exchange-correlation effects are treated using the revised Perdew-Burke-Ernzerhof functional for solids (PBEsol) \cite{PhysRevLett.100.136406} within the generalized gradient approximation (GGA) \cite{PhysRevLett.77.3865}.
The in-plane lattice constant is fixed to the value of bulk SrTiO$_3$ ($a$=$b$=3.905\,\AA), while the out-of-plane lattice constant $c$ and all atomic positions are allowed to relax. The structural optimizations are converged until the forces on atoms fall below $0.01$\,eV/{\AA}.  A $\Gamma$-centered $14 \times 14 \times 2$ $\mathbf{k}$-mesh is used for Brillouin zone sampling.

After structural relaxations, we calculate the electronic band structures using the full-potential linearized augmented plane wave (LAPW) method as implemented in \textsc{Wien2k} \cite{blaha2001wien2k}. Nd $4f$ states are treated as {\it open-core} states.
We construct Wannier functions and the consequent tight-binding Hamiltonians for the correlated orbitals of the localized states---namely Nd 5$d$, Ni 3$d$ and/or Ti 3$d$ orbitals---by projecting the DFT bands near the Fermi surface onto maximally localized Wannier functions \cite{RevModPhys.84.1419}, using \textsc{wannier90} \cite{mostofi2008wannier90,Pizzi2020}, interfaced through \textsc{wien2wannier} \cite{Kunes2010a}.
These tight-binding Hamiltonians serve as the respective non-interacting Hamiltonian in the following DMFT calculations for the two surface models shown in Fig.~\ref{fig:structure}.
We employ the inter-orbital Coulomb interaction $U'$=3.10\,eV (2.00\,eV) and Hund's exchange $J$=0.65\,eV (0.25\,eV) for Ni 3$d$ (Nd 5$d$) orbitals.

These values were calculated in Ref.~\cite{Si2020} for bulk NdNiO$_2$ within the constrained random phase approximation (cRPA) \cite{PhysRevB.77.085122}, as implemented in \textsc{Vasp} \cite{PhysRevMaterials.9.015001}.
For simplicity, we assign the same parameters to Ti 3$d$ and Sr 4$d$ orbitals as those used for Ni 3$d$ and Nd 5$d$ orbitals, respectively. As  before \cite{Si2020}, we use the Anisomov double counting \cite{Anisimov1991}.
Let us note that there 
is some uncertainty in the cRPA calculation of the interaction as well as the double counting correction.
This can slightly affect how much the Nd 5$d$ are shifted up when their interaction is included (which in any case is better than not to include this interaction).

The multi-band DMFT self-consistency equations are solved at room temperature (300\,K) by continuous-time quantum Monte-Carlo (QMC) simulations in the hybridization expansion \cite{Gull2011a} using the \textsc{W2dynamics} package \cite{w2dynamics2018}.
We perform the analytic continuation of self-energies to the real-frequency axis using the maximum entropy method \cite{PhysRevB.44.6011}, implemented in the \textsc{ana\_cont} code \cite{Kaufmann2021}.

\section{Results}
\label{Sec:results}
\subsection{DFT}
\label{Sec:DFT}

\begin{figure} [tb]
        \centering
        \includegraphics[width=\linewidth]{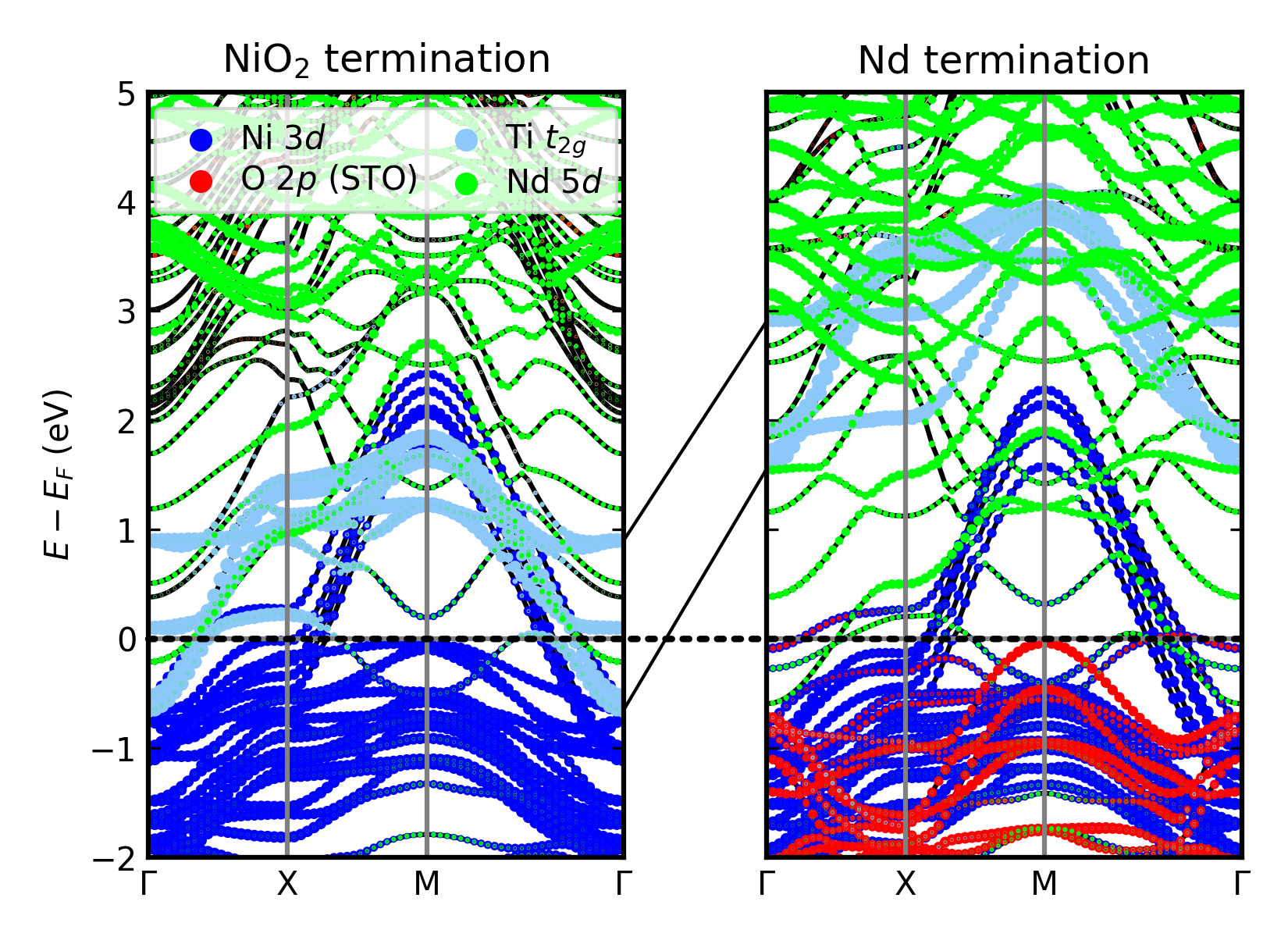}
        \caption{\label{fig:DFT_projs}  DFT  band structures for the NiO$_2$-terminated surface (left) and the Nd-terminated one (right). The band projections are colored according to their
        atomic character, as in Fig.~\ref{fig:structure}.}
\end{figure}

\begin{figure} [tb]
        \centering
        \includegraphics[width=\linewidth]{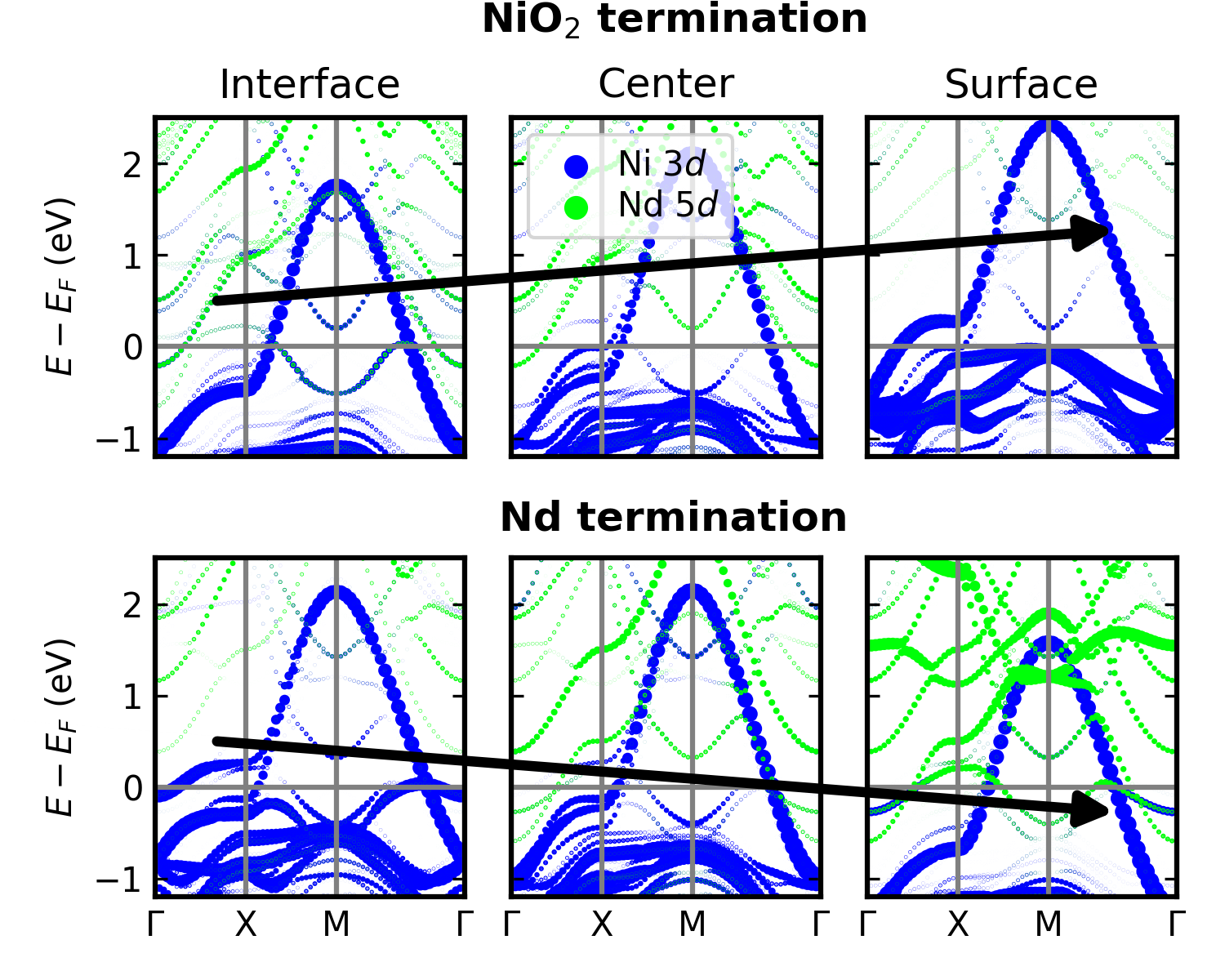}
        \caption{\label{fig:DFT} DFT band structures of the  NiO$_2$-terminated surface (top) and the Nd-terminated one (bottom) projected onto the interface, center and surface layer. Nd and Ni contributions are colored in green and blue, respectively.  }
\end{figure}

Let us begin by discussing the DFT results for the two models shown in Fig.~\ref{fig:structure}, each consisting of five NdNiO$_2$ 
layers in contact with two SrTiO$_3$ layers that represent the substrate.  
These supercell geometries can cover both electronic reconstruction and lattice deformation. Modeling more realistic surface effects, such as the adsorption of oxygen or hydrogen, incomplete layer growth, or the formation of superstructures
is beyond the scope of this work as this would require even larger cells and  experimental evidence which of these prospective surface effects are realized.

The relaxed structures in 
Fig.~\ref{fig:structure} reveal a noticeable displacement of the Ni atoms along $z$-direction in the surface and interface layer, relative to the intra-layer oxygen atoms. In contrast, the innermost layers remain almost unchanged, closely resembling the ideal planar NiO$_2$ sheets as in bulk NdNiO$_2$. At the interface, the oxygen atoms are significantly closer to the Nd$^{3+}$ ions than to the Sr$^{2+}$ ions in both models,  due to the stronger electrostatic attraction between Nd$^{3+}$ and O$^{2-}$. The NiO$_2$-terminated surface (left structure in Fig.~\ref{fig:structure}) is energetically more stable than the Nd-terminated surface (right panel in Fig.~\ref{fig:structure}) by
$E_{\text{Nd}} - E_{\text{NiO}_2}$=2.2\,eV per supercell.
This DFT finding is consistent with experiments, which predominantly report the formation of an $n$-type interface with SrTiO$_3$.

Fig.~\ref{fig:DFT_projs} presents the DFT band structures for both slab models, with the dominant atomic and orbital character of each band highlighted using the same color scheme as in Fig.~\ref{fig:structure}. 
For both slabs, the five Ni 3$d_{x^2-y^2}$ bands cross the Fermi energy.
These five bands originate from the slab geometry with five NiO$_2$ layers and are slightly offset. Additionally, one Nd-5$d$-derived band crosses the Fermi energy around the M point and one or two (in the left and right panel of Fig.~\ref{fig:DFT_projs}, respectively) around the $\Gamma$ point. 
Let us note that the A pocket, observed in bulk nickelates, corresponds for the finite slab calculation to the electron pocket at the M point of the surface Brillouin zone. Hence, while using up to this point the notation ``A'' pocket from the literature, we will use   ``M'' point henceforth, as this is the correct momentum of our finite slab. 

The electron pocket around the M point is predominately of $t_{2g}$ character close to the M point itself. This is because, without hybridization,  the Nd 5$d$ band would go down to much lower energies (-1.8\,eV; where one sees a green contribution in Fig.~\ref{fig:DFT_projs}). The band minimum of the M pocket around -0.4\,eV  instead stems from the reconstruction with the Ni $t_{2g}$ orbital (and is thus blue in  Fig.~\ref{fig:DFT_projs}).

In the case of the NiO$_2$ termination (left panel of  Fig.~\ref{fig:DFT_projs})
additionally the Ti $t_{2g}$ bands (light blue) cross the Fermi energy.
To gain further insight, it is helpful to examine the layer-resolved band structure shown in Fig.~\ref{fig:DFT}, focusing on the interface, center, and surface layer. Those plots reveal a pronounced layer dependence and a noticeable shift in the band centers, as indicated by the arrows. This shift arises from the internal polar field along the $z$-direction, which acts in opposite directions for the two slabs calculated.
For the NiO$_2$-termination, the  negatively charged (NiO$_2$)$^{3-}$ layer only has a compensating  Nd$^{3+}$ layer on one side. This puts the  negatively charged electrons to a higher potential energy; their bands are shifted upward. 
In more detail, the Ni bands shift progressively upward in energy when moving from the interface toward the surface. This upward shift leads to a depletion of Ni electrons at the surface, with the excess charge transferred into the Ti bands of the substrate, which then become partially occupied. As shown in Fig.~\ref{fig:DFT_projs} (left panel), the Ti 3$d$ orbitals at the interface clearly cross the Fermi level, indicating a transfer of electrons from surface Ni 3$d$ to interface Ti 3$d$ states. This electronic reconstruction helps to compensate the internal polar electric field generated by the 
alternating (NiO$_2$)$^{3-}$ and Nd$^{3+}$ layers, thereby contributing to the stabilization of the NiO$_2$-terminated surface. 

For the Nd-terminated surface, the polar field, as indicated in Fig.~\ref{fig:DFT} (bottom), points in the opposite direction compared to the NiO$_2$-terminated surface.
Consequently, the interfacial Ti bands are located well above the Fermi level, as seen in Fig.~\ref{fig:DFT} (right), and electrons accumulate at the surface.

In the following DMFT calculations, it is therefore necessary to include the electron-doped Ti bands for the NiO$_2$-terminated surface, while this is not necessary  for the Nd-terminated one.
Accordingly, we project the corresponding DFT bands to 75 Wannier functions in the former case, and 55 in the later. These Wannier functions include the Ni 3$d$, Nd 5$d$, and interstitial $s$ orbitals;  as well as the Ti 3$d$ and Sr $5d$
for the NiO$_2$-terminated surface.
Details of the orbitals and energy window selected to target DFT bands, along with a comparison of the DFT and Wannier bands, are provided in Section~S1 in SM \cite{SM}. Finally, as we will demonstrate later in the DMFT results (see Figs.~\ref{fig:Awk_Nd_NiO2} and \ref{fig:Awk_Nd_Nd}), the effects of the polar field observed in the DFT band structure persist even when electronic correlations are included via DMFT.

\subsection{DMFT: NiO$_2$ termination}
\label{Sec:DMFTNiO2}
  \begin{figure} [tb]
        \includegraphics[width=1\linewidth]{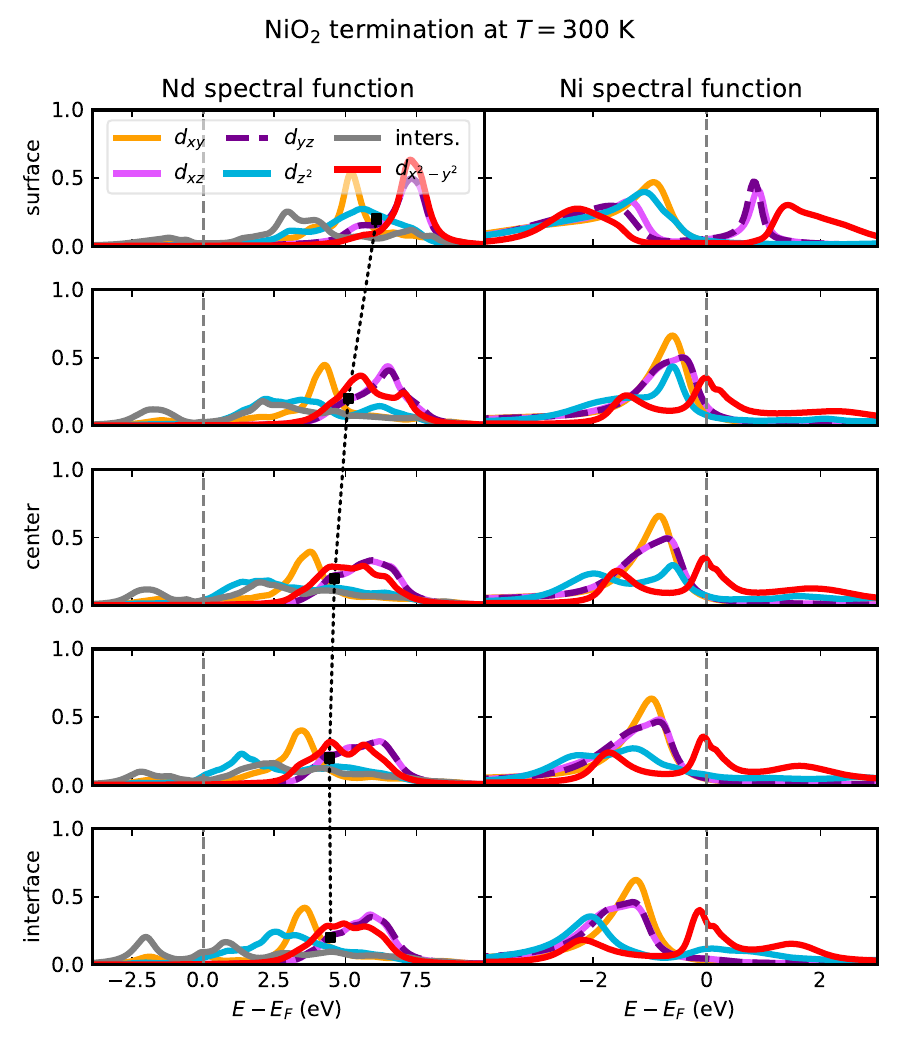}
        \caption{\label{fig:Aw_NiO2}Layer-resolved, $\mathbf{k}$-integrated spectral function of Nd 5$d$ orbitals (left column) and Ni 3$d$ orbitals (right column) for the NiO$_2$-terminated slab. The gray line is the additional interstitial $s$ orbital, which is located within the Nd planes. The bottom row is at the interface, the top row is at the surface. The black dotted line visualizes the polar field shift; the vertical dashed line indicates the Fermi energy. }
    \end{figure}
    
\begin{figure} [tb]
        \includegraphics[width=1\linewidth]{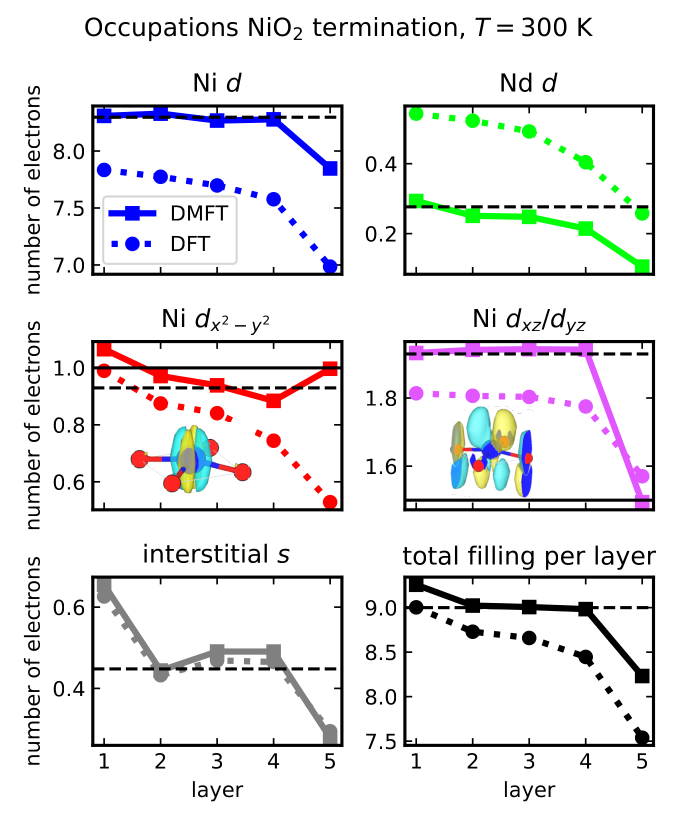}
        \caption{\label{fig:filling_NiO2}Comparison of filling in DMFT (solid lines, squares) and DFT (dotted lines, circles) for different orbitals in case of the NiO$_2$-terminated slab. Layer 1 is at the interface, layer 5 is at the surface. Dashed, black horizontals are the fillings for bulk NdNiO$_2$ in DMFT. Black, solid lines in the central row indicate half-filling of Ni $3d_{x^2-y^2}$ and three-quarter filling of the degenerate Ni $3d_{xy}/3d_{yz}$ orbitals, respectively. The Mott-insulating surface Ni $3d_{x^2-y^2}$ and $3d_{xz}$ Wannier functions are depicted in the two central plots.}
    \end{figure}

Let us now turn to the DMFT results, starting with the NiO$_2$ termination of Fig.~\ref{fig:structure} (left).
Fig.~\ref{fig:Aw_NiO2} presents the ${\mathbf k}$-integrated DMFT spectral functions, resolved for layer, atom, and orbital.
The polar field is most clearly seen in the unoccupied Nd bands. Their center of gravity shifts, as indicated by the dashed line in Fig.~\ref{fig:Aw_NiO2}, from layer to layer, with the surface layer at the highest energies.
Correspondingly, the filling of the Nd 5$d$ states in Fig.~\ref{fig:filling_NiO2} decreases toward the surface. Note that the Nd 5$d$ filling stems from two effects:
(i) the M and $\Gamma$ pocket where the Nd bands cross the Fermi energy and (ii) the hybridization with the Ni 3$d$ states. Both effects are reduced with the gradual up-shift of the Nd bands when approaching the surface.

Including dynamical correlation effects via DFT+DMFT leads to significantly more pronounced modifications in the Ni 3$d$ states. The most striking difference is that, while all layers remain metallic in DFT (Fig.~\ref{fig:DFT}), the NiO$_2$ surface layer becomes a Mott insulator in DFT+DMFT. This is clearly evidenced by the DMFT spectral function $A(\omega)$ shown in Fig.~\ref{fig:Aw_NiO2}, where a well-defined gap appears at the Fermi level for the Ni 3$d$ states in the surface layer (topmost panel).

A physical explanation of the origin of this Mott insulating state can be gained from the orbitally resolved filling of the different states
which is shown in  Fig.~\ref{fig:filling_NiO2}. The total DFT and DMFT fillings of the  Nd 5$d$ and Ni 3$d$ generally follow the trend dictated by  the polar field.
This means that, at the surface, the  Nd 5$d$ and Ni 3$d$ states exhibit reduced fillings. However, the emergence of a Mott-insulating state at the surface is linked to an abrupt change in the DMFT filling. In particular, against the trend of the polar field, the Ni $d_{x^2-y^2}$ orbital becomes approximately half-filled, with an again larger occupation close to one electron per site.
The overall reduction in Ni 3$d$ electron occupancy (Fig.~\ref{fig:filling_NiO2}, top-left)  is realized through the depopulation of the Ni $d_{xz}$ and $d_{yz}$ orbitals, both of which have fillings around 1.5 (Fig.~\ref{fig:filling_NiO2}, center-right; i.e., half a hole per orbital). These orbitals are more sensitive to surface effects because their orbitals' lobes extend more along the $z$-direction, making them more sensitive to reduced coordination and screening effects. This presence of approximately two holes across three Ni 3$d$ orbitals favors the formation of a spin-1 local moment and stabilizes the Mott-insulating state at the surface.

\begin{figure*} 
        \includegraphics[width=1\linewidth]{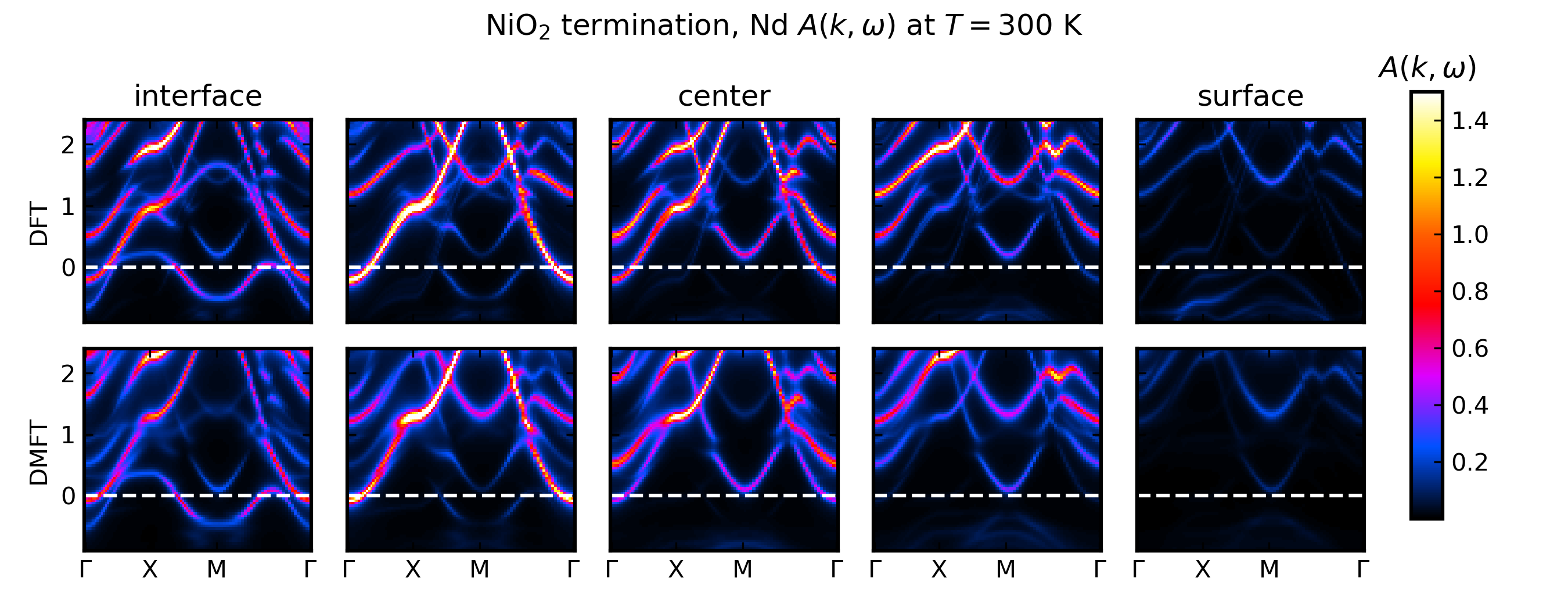}
        \caption{\label{fig:Awk_Nd_NiO2}Layer-resolved, $\mathbf{k}$-dependent spectral function of Nd $5d$ orbitals in the NiO$_2$-terminated surface obtained from DFT (upper row) and DMFT (bottom row) along a high-symmetry path through the 2D Brillouin zone. Left column is at the interface, right column is at the surface. }
    \end{figure*}

Let us note that without the surface layer, i.e., in a (SrTiO$_3$)$_4$$|$(PrNiO$_2$)$_6$ superlattice,
the NiO$_2$ surface (now interface)  layer is not  
 Mott insulating  any longer \cite{Wenfeng_private}.
Let us further note that in Fig.~\ref{fig:filling_NiO2} (top-left panel), the Ni 3$d$ electron filling is not exactly 8 electrons per Ni site, as one might expect for a two-hole configuration. This discrepancy arises from how electron are counted for a hybridized multi-orbital system in DFT and DMFT. Similar to the situation in bulk NdNiO$_2$ (dashed black line), the Ni 3$d$ orbitals in our surface model  strongly hybridized with the Nd 5$d$ orbitals. As a result, the states near and below the Fermi level---although predominantly of Ni 3$d$ character---also contain significant Nd 5$d$ admixture (see Fig.~\ref{fig:Aw_NiO2}).
Conversely, there is also some Ni 3$d$ contribution to the unoccupied Nd 5$d$ states above the Fermi level. Due to this hybridization, the effective low-energy degrees of freedom resemble a 3$d^8$ configuration, even though the occupation of the ``pure'' Ni 3$d$ orbitals in the Wannier basis used for the DFT+DMFT calculations is somewhat lower. 
A qualitatively similar reduced   $d_{x^2-y^2}$  occupation has also been found in the 10-band model without interstitial $s$ orbtial \cite{Kitatani2020}. For determining whether there is genuine multi-orbital physics, one should thus always have a look at the spectral function.  Fig.~\ref{fig:Aw_NiO2} clearly shows an upper Hubbard band for Ni $d_{x^2-y^2}$, $d_{xz}$ and $d_{yz}$ orbitals for the surface layer, i.e., multi-orbital physics. 
In contrast, for the central layers,  only the $d_{x^2-y^2}$ orbital forms spectral weight  above the Fermi energy (beyond the aforementioned hybridization with the Nd orbitals). One has single orbital physics. Concomitantly, the $\mathbf k$-resolved spectral function in Fig.~\ref{fig:Awk_Nd_NiO2} (discussed in the next paragraph) shows only the Ni $d_{x^2-y^2}$ orbital (plus M and $\Gamma$ pockets) but no other Ni orbital crossing the Fermi energy.

\begin{figure*} 
        \includegraphics[width=1\linewidth]{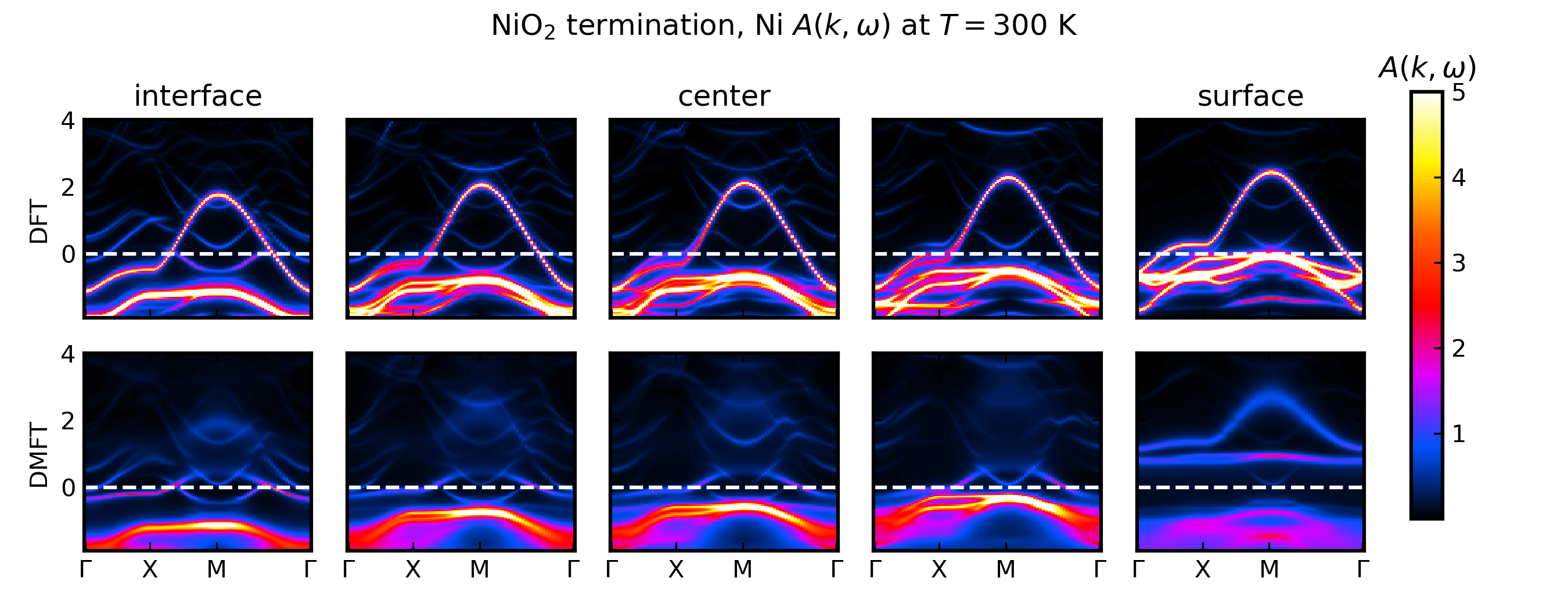}
        \caption{\label{fig:Awk_Ni_NiO2}Same as  Fig.~\ref{fig:Awk_Nd_NiO2}, but now projected onto Ni $3d$ orbitals.}
    \end{figure*}

To gain deeper insight into the electronic structure of the surface model, we compute layer-resolved DMFT spectral functions $A(\mathbf{k}, \omega)$ for Nd 5$d$ (Fig.~\ref{fig:Awk_Nd_NiO2}) and Ni 3$d$ states (Fig.~\ref{fig:Awk_Ni_NiO2}). This analysis is motivated by the fact that Nd 5$d$ and Ni 3$d$ states both contribute to the Fermi surface, as supported by recent ARPES measurements~\cite{Ding2024,Sun2025}.
The spectral functions for the interstitial $s$ orbitals and Ti $3d$ states are provided in the Supplemental Material (Section S2) \cite{SM}. In all figures (Figs.~\ref{fig:Awk_Nd_NiO2}-\ref{fig:Awk_Ni_NiO2}), DFT results are shown in the top panels, while the corresponding DMFT spectra are displayed in the bottom panels.
In Fig.~\ref{fig:Awk_Nd_NiO2}, the Nd-derived bands exhibit a clear layer dependence, shifting to higher energies as one moves toward the surface. This is a direct consequence of the built-in polar electric field. Notably, in the surface and subsurface layers, the electron pocket at the $\Gamma$-point---clearly present in bulk NdNiO$_2$ \cite{Kitatani2020}---disappears. This pocket reemerges only deeper in the slab, from the bulk-like central layers down to the interface, reflecting the impact of surface polarity and dimensional confinement on the band structure.

The influence of the polar field is similarly evident in the Ni-derived bands shown in Fig.~\ref{fig:Awk_Ni_NiO2}, for both DFT and DMFT results. Most notably, the Ni $d_{x^2-y^2}$ orbital exhibits significant renormalization. In the surface layer, this orbital even becomes Mott-insulating, consistent with the $\mathbf{k}$-integrated DMFT spectral function $A$($\omega$) discussed earlier. Additionally, as seen more clearly in the orbital-resolved spectral function (Fig.~\ref{fig:Aw_NiO2}), the Ni $d_{xz}$ and $d_{yz}$ orbitals undergo a pronounced splitting into lower and upper Hubbard bands, further underscoring the strong correlation effects near the surface.

\begin{figure*} 
        \includegraphics[width=1\linewidth]{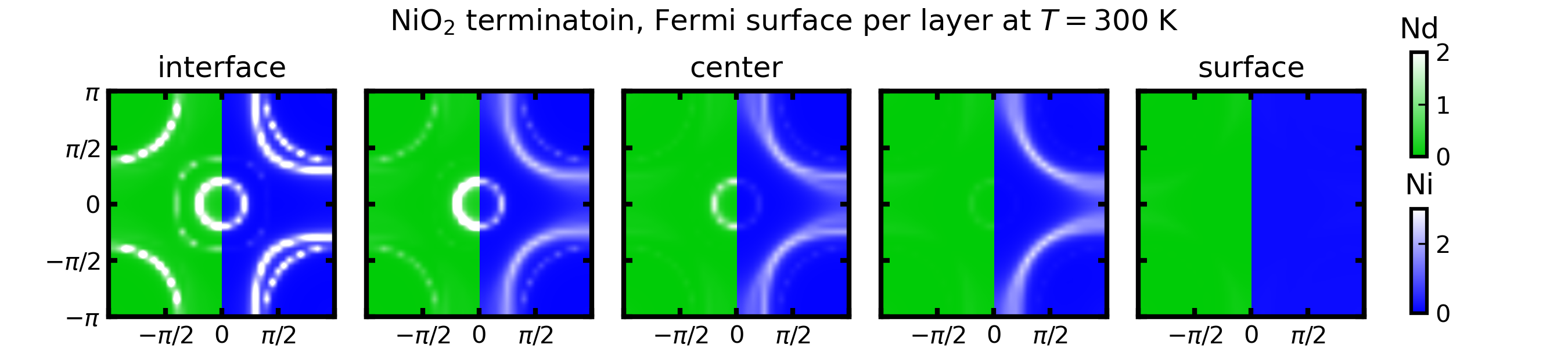}
        \caption{\label{fig:FS_NiO2}2D Fermi surface contribution of each layer and atom (green: Nd, blue: Ni). Interfacial Nd is in contact with a TiO$_2$ layer from the substrate (left end) and a NiO$_2$ layer faces to the vacuum (right end).}
    \end{figure*}

To enable a direct comparison between our simulations and experimental observations,
Fig.~\ref{fig:FS_NiO2} presents the layer-resolved DMFT Fermi surface $A({\mathbf k},\omega=0)$ for both Nd 5$d$ and Ni 3$d$ orbitals. The panels are ordered from right to left, beginning with the NiO$_2$ surface layer (labeled as layer 5) to Nd interface layer (layer 1).
Fig.~\ref{fig:FS_NiO2} clearly illustrates the Mott-insulating nature of the surface NiO$_2$ layer and the absence of the $\Gamma$-centered electron pocket in both the surface (layer 5) and subsurface (layer 4).
In contrast, the $\Gamma$ pocket emerges in the deeper layers, particularly near the interface and central bulk-like regions \cite{noteGamma}. 
Further, the symmetry breaking at the interface, due to the displacement of the Ni atom and the loss of the original $P4/mmm$ symmetry, enhances the hybridization between Ni 3$d$ and Nd 5$d$ orbitals. As a result, the $\Gamma$ pocket appears not only in the Nd-derived spectral weight but also becomes visible in the Ni spectrum. Conversely, the cuprate-like hole Fermi surface originating from the Ni $d_{x^2-y^2}$ orbital is also reflected in the Nd states due to this orbital mixing.

The employed 5-layer NdNiO$_2$ model corresponds to a 2\,nm thick film in experimental epitaxial growth. An important question is whether this thickness is sufficient to reliably capture surface and interface effects. To address this, we present additional tests in Section~S3 of the Supplemental Material \cite{SM}, demonstrating that the key results remain robust when increasing the slab thickness or lowering the temperature. Furthermore, even in calculations without the SrTiO$_3$ substrate, the NiO$_2$ surface layer remains insulating. However, when using a simplified tight-binding Hamiltonian that includes the polar field and truncates hopping at the vacuum boundary---but omits structural relaxations and the resulting crystal field modifications at the surface---the NiO$_2$ surface becomes metallic. This underscores the critical role of lattice relaxation and surface crystal field effects in stabilizing the Mott-insulating state, as demonstrated in SM~\cite{SM} Section~S3.

\subsection{DMFT: Nd termination}
\label{Sec:DMFTNd}

\begin{figure} [tb]
        \includegraphics[width=1\linewidth]{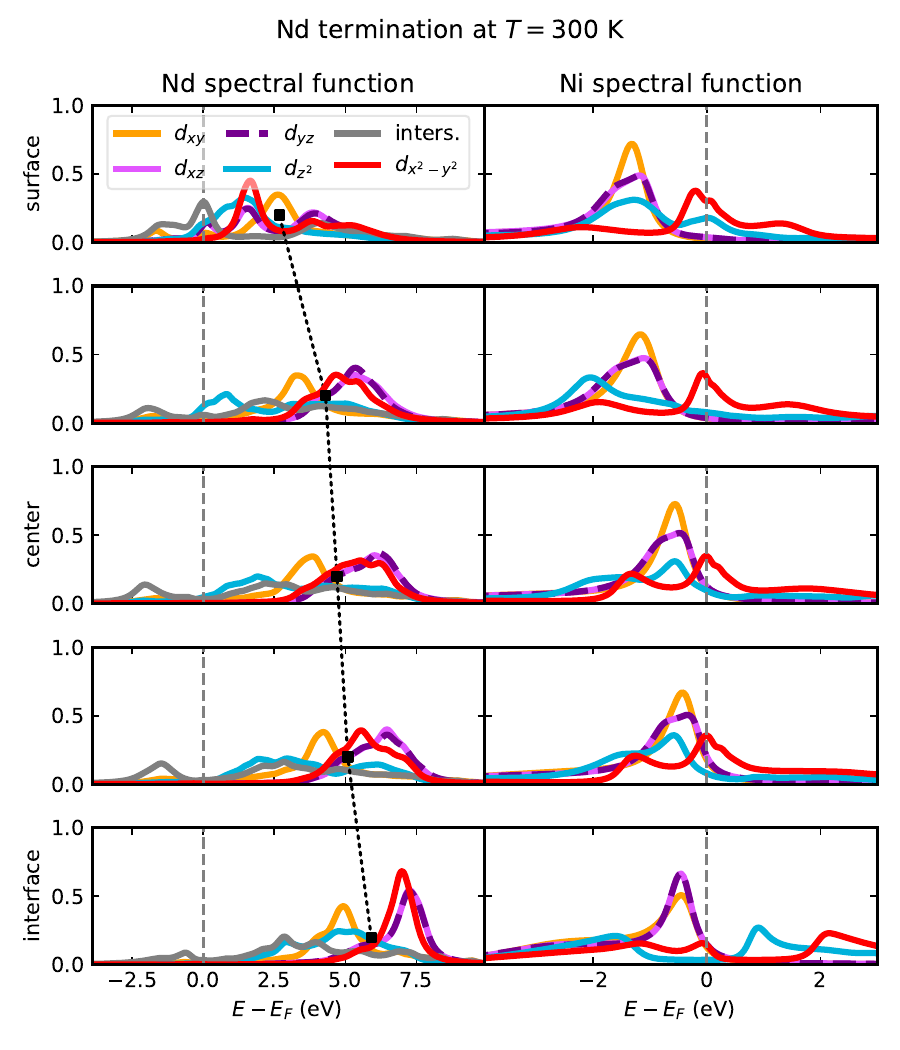}
        \caption{\label{fig:Aw_Nd}Same as Fig.~\ref{fig:Aw_NiO2}, but for the Nd-terminated surface.}
    \end{figure}

\begin{figure}  [tb]
        \includegraphics[width=1\linewidth]{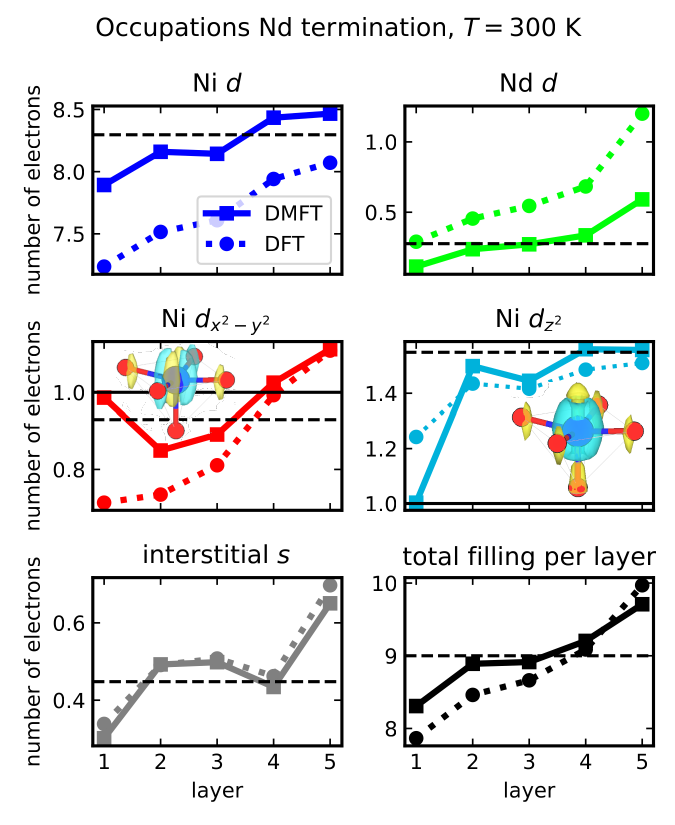}
        \caption{\label{fig:filling_Nd}Same as  Fig.~\ref{fig:filling_NiO2}, but for the Nd-terminated surface. Here, the center-right panel shows the filling of Ni $3d_{z^2}$ orbitals, which is close to half filling at the interface.}
    \end{figure}

Next, we turn to the Nd-terminated surface, shown in the right panel of Fig.~\ref{fig:structure}. In this case, the intrinsic polar field, as indicated by the dashed line in Fig.~\ref{fig:Aw_Nd}, points in the opposite direction compared to the NiO$_2$-terminated structure, placing the surface Nd bands at the lowest energy. As a result, the Nd 5$d$ orbitals contribute significantly at the Fermi level in DMFT calculations. The Ni surface layer remains metallic for this surface termination, as evidenced by the absence of a band gap in the spectral function $A$($\omega$) (see Fig.~\ref{fig:Aw_Nd}, top-right panel).
This reversed polar field also leads to an opposite trend in both total and Ni $d$ orbital fillings compared to the NiO$_2$-terminated case, as shown in Fig.~\ref{fig:filling_Nd}. Again, we observe an abrupt change in the occupation of the Ni $d_{x^2-y^2}$ orbital.  However, for Nd-terminated surface model, it happens in the interface layer (layer 1), which becomes half-filled (see Fig.~\ref{fig:filling_Nd}, center-left panel). The reduced Ni 3$d$ occupancy here originates from the $d_{z^2}$ orbital, which is also approximately half-filled, as seen in the center-right panel of Fig.~\ref{fig:filling_Nd}.
Thus, similar to the surface layer in the NiO$_2$-terminated structure, the interface layer in the Nd-terminated case hosts approximately two holes in the Ni 3$d$ orbitals, leading to the emergence of an orbitally selective Mott-insulating state at the interface (see right-bottom panel in Fig.~\ref{fig:Aw_Nd}). Here, there is a Mott-Hubbard gap for the Ni $d_{z^2}$  orbital, while the
Ni $d_{x^2-y^2}$ still shows a quasiparticle peak at the Fermi energy (zero).
That is, both surface terminations exhibit similar physics: the layer with the lowest electrostatic potential (the surface layer for NiO$_2$-termination and the interface layer for Nd termination) tends to localize holes and develop a (partial) Mott-insulating behavior. Additional holes are hence no longer accommodated in the $d_{x^2-y^2}$ orbital alone but instead occupy other Ni orbitals. This signals the emergence of multi-orbital behavior \cite{kang2023infinite,PhysRevB.102.161118,PhysRevLett.124.207004}.

For the NiO$_2$-terminated surface, which features a square planar crystal field, the additional holes are primarily located at the Ni $d_{xz}$ and $d_{yz}$ orbitals. In contrast, the interface layer of the Nd-terminated surface exhibits a NiO$_5$ pyramidal crystal field. Here, the Ni $d_{z^2}$ orbital, that points toward the additional (negatively charged) apical oxygen, is pushed up in energy and becomes the first to be depopulated after the $d_{x^2-y^2}$ orbital. This difference in oxygen coordination thus accounts for the distinct orbital characters involved in the emergent multi-orbital physics of the two terminations.

\begin{figure*}  [tb]
        \includegraphics[width=1\linewidth]{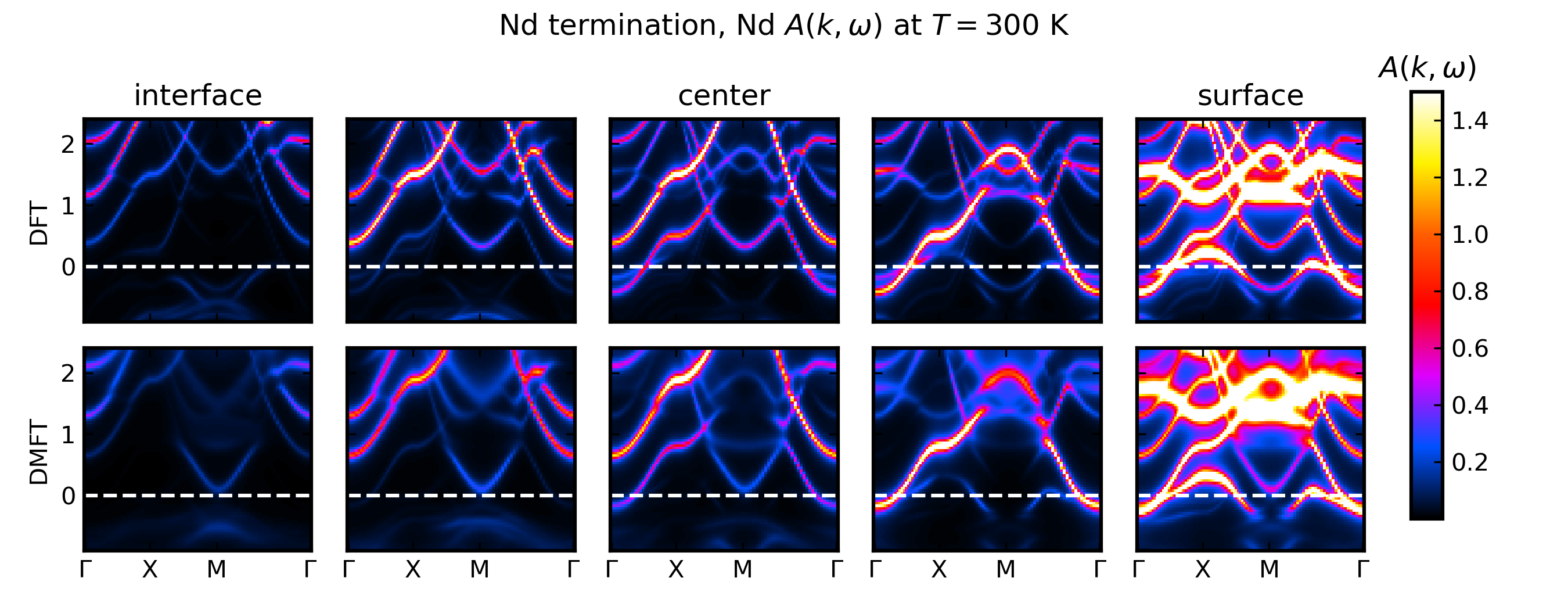}
        \caption{\label{fig:Awk_Nd_Nd}Same as Fig.~\ref{fig:Awk_Nd_NiO2}, but for the Nd-terminated surface.}
    \end{figure*}

\begin{figure*}  [tb]
        \includegraphics[width=1\linewidth]{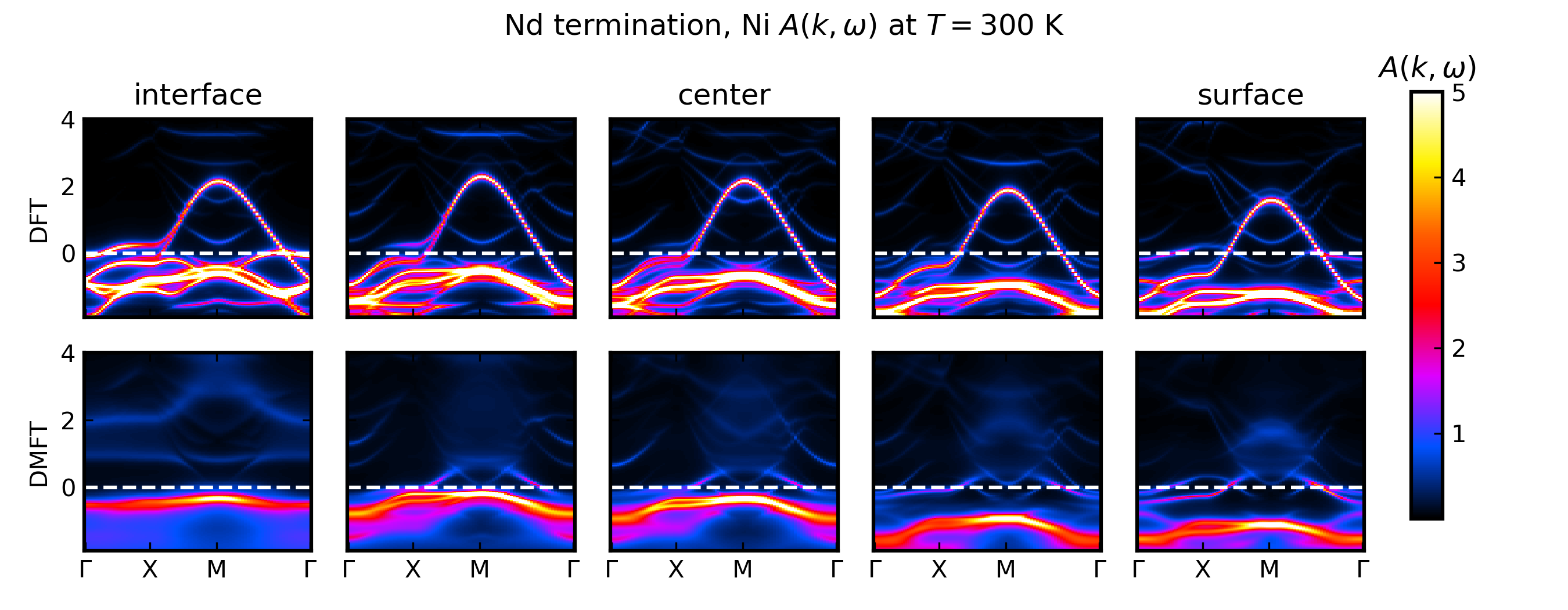}
        \caption{\label{fig:Awk_Ni_Nd}Same as Fig.~\ref{fig:Awk_Nd_NiO2}, but for the Nd-terminated surface, and projected onto Ni $3d$ orbitals.}
    \end{figure*}

\begin{figure*}  [tb]
        \includegraphics[width=1\linewidth]{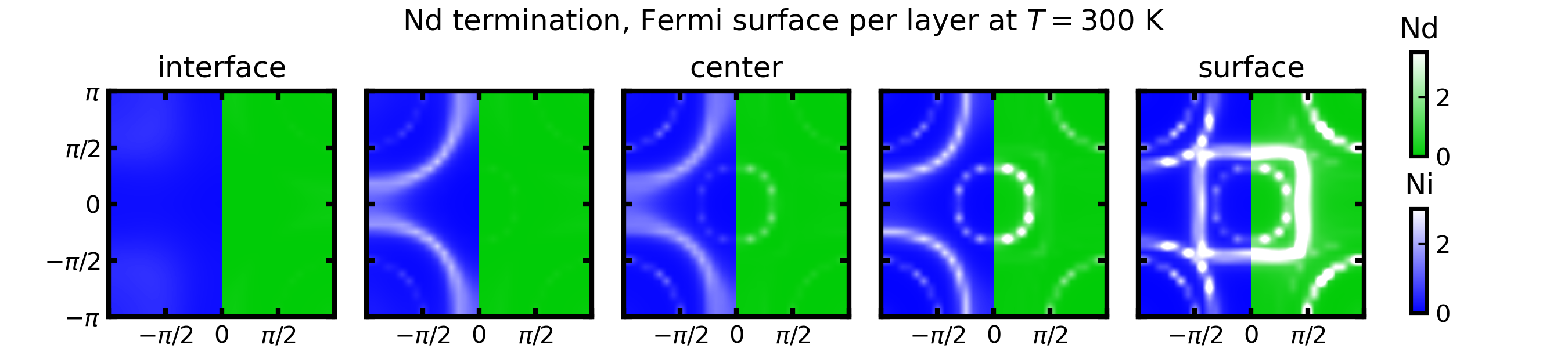}
        \caption{\label{fig:FS_Nd}Same as Fig.~\ref{fig:Awk_Nd_NiO2}, but for the Nd-terminated surface. Note that here interfacial NiO$_2$ is in contact with a SrO layer of the substrate (left end) and Nd faces to the vacuum (right end).}
    \end{figure*}

The reversed polar field in the Nd-terminated surface is also evident in the layer- and atom-resolved spectral functions of the Nd 5$d$ and Ni 3$d$ states, shown in Fig.~\ref{fig:Awk_Nd_Nd} and Fig.~\ref{fig:Awk_Ni_Nd}, respectively. Similar as in the other calculations in this paper, the Ni 3$d$ orbitals exhibit strong renormalization in DMFT, while the Nd 5$d$ bands are primarily affected by a layer-dependent shift. Notably, this shift results in prominent electron pockets at the $M$ and $\Gamma$ points, as seen in the rightmost panel of Fig.~\ref{fig:FS_Nd}. In contrast, these Nd-derived pockets are absent at the interface layer, underscoring the spatial variation in electronic structure induced by the polar field.

\section{Conclusion}
\label{Sec:conclusion}

In this work, we have investigated the surface and interface electronic structure of the infinite-layer nickelate superconductor NdNiO$_2$ using DFT combined with DMFT. By constructing two different realistic slab models we have uncovered a strong layer-dependent modulation of electronic correlations, band filling, and orbital polarization. For lack of more detailed experimental surface information, we have studied an ideal NiO$_2$- and Nd-terminated surface, without surface reconstruction.
The influence of the intrinsic polar electric field is inherent to both the surface and interface. Particularly strong effects are found for  the outermost surface and interface layers.

Especially, our DFT+DMFT calculations show that the NiO$_2$-terminated surface undergoes a Mott transition in the topmost NiO$_2$ layer, which becomes insulating. This behavior is not captured at the DFT(+U) level and highlights the crucial role of local electronic correlations and methods beyond DFT, such as DMFT. The Mott-insulating state originates from the switch from a 3$d^9$ to a 3$d^8$
electronic configuration at the surface. Here, the $d_{x^2-y^2}$ orbital becomes nearly half-filled,  while the two $d_{xz/yz}$ orbitals accommodate the remaining holes. This turns the physics from  single-Ni-orbital physics  to multi-orbital physics of the surface layer.
The redistribution of electrons is driven not only by the layer-dependent polar field but also by the modified crystal field environment at the surface.

In contrast, the Nd-terminated surface displays a reversed polar field, leading to different electronic reconstruction compared with NiO$_2$-terminated surface. Here, it is the interface NiO$_2$ layer (rather than the surface layer, as in NiO$_2$-terminated surface) that becomes 
3$d^8$ and an orbital-selective Mott insulator with the half-filled Ni $d_{z^2}$ band being insulating and the half-filled  Ni $d_{x^2-y^2}$ band being metallic. Let us note that this 
orbital-selective Mott insulator is close to becoming a full Mott insulator.
For both terminations the layer with the lowest electrostatic potential is (partially) insulating. The nature of the participating additional Ni orbitals differs due to the local ligand geometry---square planar versus square pyramidal---highlighting the sensitivity of orbital occupation and correlation strength to local coordination.

Our analysis of the DMFT spectral function,  momentum-, layer- and orbital-resolved, reveals substantial band renormalization and the emergence or disappearance of Fermi surface features such as the $\Gamma$ and $M$ pockets depending on layer depth and surface termination. 
From a methodological standpoint, we have demonstrated the necessity of incorporating (dynamical) correlations, spatial inhomogeneity, and realistic structural relaxations to accurately capture the electronic behavior of the surfaces and interfaces of nickelate superconductors. Our findings have broad implications for understanding the emergent phenomena in nickelate heterostructures and thin films. One can speculate that the observed layer-dependent doping through a polar field  might also play a role for the recently observed superconductivity in undoped PrNiO$_2$|SrTiO$_3$ superlattices. 

{\it Note added.} While finalizing this manuscript, measurements of ARPES spectra for NdNiO$_2$
became available \cite{li2025} and show no $\Gamma$ pocket.

\section{Acknowledgments.}
We thank Viktor Christiansson, Eric Jacob, Wenfeng Wu, and Eduard Schröder for helpful discussions.
We further acknowledge funding by the Austrian Science Funds (FWF) through project DOI 10.55776/I5398. The DFT+DMFT calculations have been mainly done on the Vienna Scientific Cluster (VSC).
L.~S.~acknowledges support from the National Natural Science Foundation of China (Grant No.~12422407).

For the purpose of open access, the authors have applied a CC BY public copyright license to any Author Accepted Manuscript version arising from this submission.

\section{Data availability}
The input and output data for the calculations presented in the main text are openly available \cite{nomad_repo}.

\clearpage
\bibliography{main}

\pagebreak
\widetext
\newpage
\begin{center}
    \textbf{
    \huge
    Supplementary material for ``Surfaces and interfaces of infinite-layer nickelates studied by dynamical mean-field theory''
    }
\end{center}
\vspace{1cm}
This supplemental material contains details and additional calculations for the two slab models discussed in the main text. It is organized as follows:\\ 
Section~\ref{sec:wannierization} gives further information about the Wannierization procedure, used to create the non-interacting Hamiltonians. 
Section~\ref{sec:add_figs} provides additional figures for the calculations of the (A) NiO$_2$-terminated and (B) Nd-terminated surface presented in the main text. They include the DMFT self energy, the total and interstitial $s$ spectral function, and the full Fermi surface.\\
Section~\ref{sec:other_slabs} explores the robustness of our results through calculations performed under modified conditions:
 We start with a larger slab in Section~\ref{sec:7layers} and a freestanding film (no SrTiO$_3$ substrate) in Section~\ref{Sec:NoSTO}. In Section~\ref{sec:different_T} we calculate the surfaces from the main text at lower temperatures. Finally, in Section~\ref{sec:tight_binding}, we show results for a slab tight binding Hamiltonian derived from the bulk Hamiltonian with truncated $z$-hopping to mimic a pristine surface without ionic displacements.

\renewcommand{\thefigure}{S\arabic{figure}}
\renewcommand{\thesection}{S\arabic{section}}
\renewcommand{\thetable}{S\Roman{table}}
\renewcommand{\theequation}{S\arabic{equation}}

\setcounter{equation}{0}
\setcounter{figure}{0}
\setcounter{table}{0}
\setcounter{section}{0}
\makeatletter
\renewcommand{\figurename}{Fig.}
\renewcommand{\thefigure}{S\arabic{figure}}
\renewcommand{\theequation}{S\arabic{equation}}
\renewcommand{\tablename}{Tab.}
\renewcommand{\thetable}{S\arabic{table}}

\newpage

\section{Wannierization}
\label{sec:wannierization}
\begin{figure*} [tb]
        \centering
        \includegraphics[width=\linewidth]{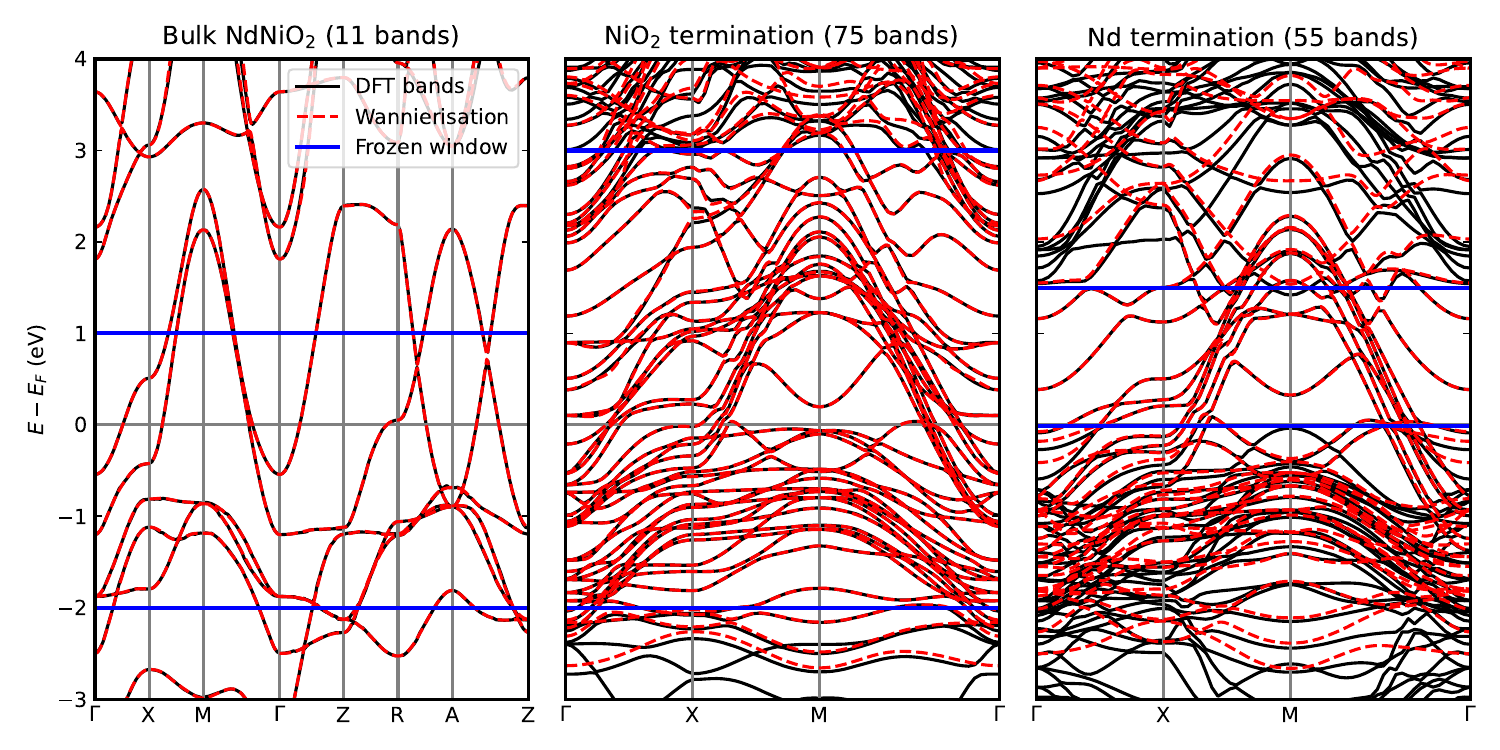}
        \caption{\label{fig:DFT_bands} Band structure of bulk NdNiO$_2$ (left), NiO$_2$-terminated (center) and Nd-terminated slab (right); see the main text Fig.~1 for the slab geometry. Black are DFT bands, red-dashed are Wannier interpolated bands. Blue horizontals indicate the respective frozen windows in \textsc{wannier90}.}
\end{figure*}

\begin{figure*} [tb]
        \centering
        \includegraphics[width=\linewidth]{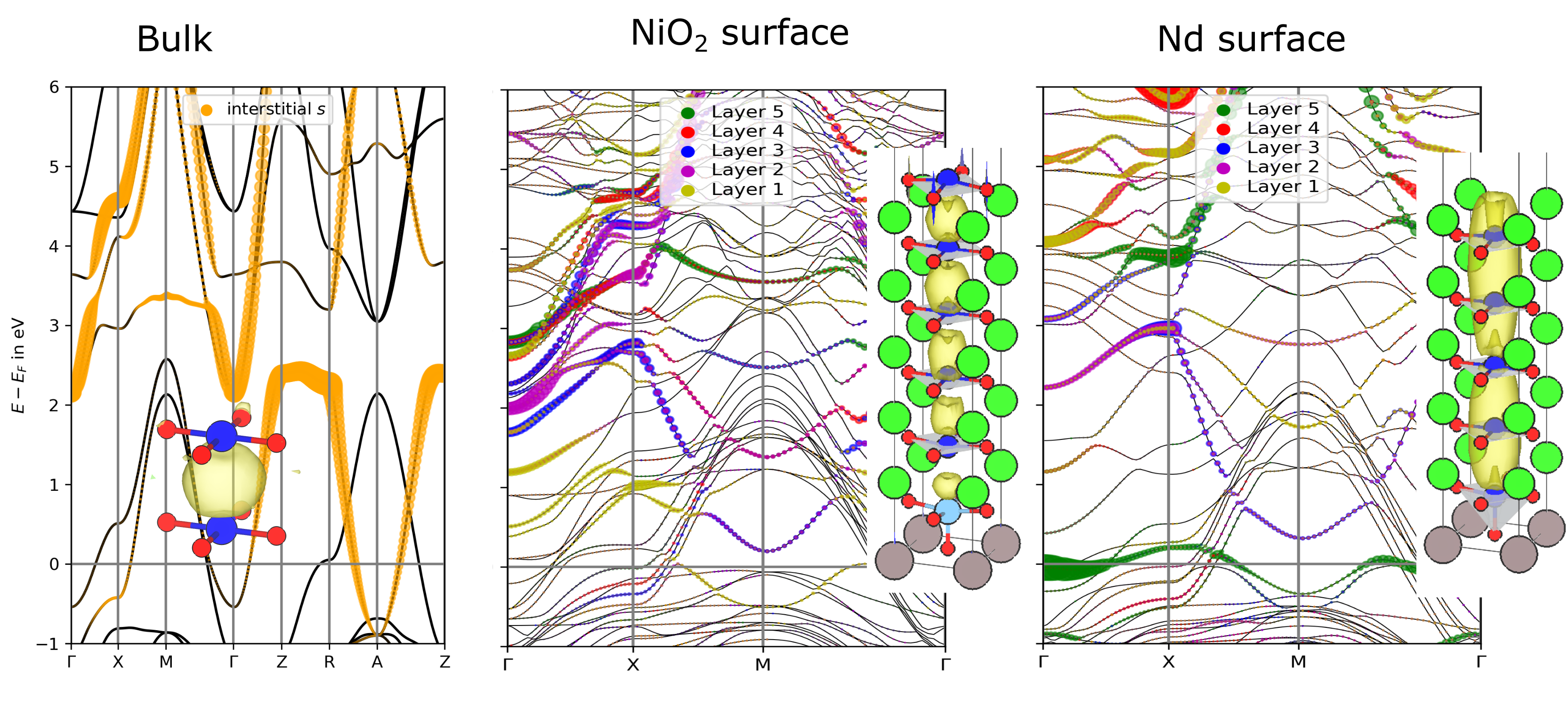}
        \caption{\label{fig:interstitial} 
        Wannier bands of bulk NdNiO$_2$ (left) and the two slabs. Shown are the projections onto the interstitial $s$ orbitals with color code  according to the Nd planes they are in. Layer 1 is at the interface, layer 5 is the surface. The insets of the figures show the sum of all interstitial $s$ Wannier functions in their respective real-space structures.}
\end{figure*}

The DFT band structure of both slabs from the main text (Fig.~1) is shown next to the bulk band structure of NdNiO$_2$ in Fig.~\ref{fig:DFT_bands}. Wannier interpolated bands, according to the orbitals listed in Table~\ref{tab:wannier}, are superimposed in red. As discussed in the main text, the polar field points in different directions for the two surfaces: The Nd-terminated surface exhibits a $p$-type interface, shifting the interfacial Ti bands to higher energies, whereas the $n$-type interface appearing for the NiO$_2$-terminated surface has the opposite effect.
This shift leads to Ti $t_{2g}$ bands crossing the Fermi level. These Ti bands are doped with electrons transferred from the NiO$_2$ surface. The Wannierization for the NiO$_2$-terminated slab, therefore, needs to include additional bands that originate not only from the NdNiO$_2$ film but also from the substrate. Although it would be enough to just include the six doped Ti $t_{2g}$ bands, for completeness we performed the projection onto all 20 substrate $d$ orbitals, resulting in 75 Wannier orbitals for the NiO$_2$-terminated slab, see Table~\ref{tab:wannierorbitals}.  \\
The interstitial $s$ orbital\cite{Gu2020}, has sometimes been neglected in low-energy models of bulk NdNiO$_2$ \cite{Si2020} as  it becomes relevant at energies of about 2\,eV, as shown in Fig.~\ref{fig:interstitial}. Due to the polar field in our slabs, however, these energy scales need to be considered as well. We thus include the interstitial $s$ orbital in our Wannier projection.

\begin{table}[tb]
\caption{\label{tab:wannier}Wannier orbitals of bulk NdNiO$_2$ and the two slab models from the main text Fig.~1.
\label{tab:wannierorbitals}}
\centering
\begin{tabular}{|c|c|c|}\hline
    system  & number of   & projections \\
    & Wannier orbitals   & \\ \hline
     bulk   & 11                    & 5 Ni $3d$\\ 
            &                       & 5 Nd $5d$ \\
            &                       & 1 interstitial $s$ \\ \hline
     NiO$_2$-termination        & 75                    & 25 Ni $3d$\\ 
        (5 layers NdNiO$_2$,    &                       & 25 Nd $5d$ \\
         2 layers SrTiO$_3$)   &                       & 10 Sr $4d$ \\
            &                   & 10 Ti $3d$ \\
            &                   & 5 interstitial $s$ \\ \hline
    Nd-termination              & 55                    & 25 Ni $3d$\\ 
        (5 layers NdNiO$_2$,    &                       & 25 Nd $5d$ \\
         2 layers SrTiO$_3$)   &                       & 5 interstitial $s$ \\ \hline
\end{tabular}
\end{table}

\clearpage 

\section{Additional figures for the two slabs of the main text}
\label{sec:add_figs}
Here, we show more data for the two slab systems with different surface terminations discussed in the main text. See Fig.~1 of the main text for the two slabs. The DMFT calculations in this section and in the main text are at room temperature (300\,K).   

\subsection{NiO$_2$-terminated slab}
\label{sec:add_figs_nio2}

\begin{figure}[tb]
     \centering
     \begin{subfigure}[b]{\textwidth}
         \centering
         \includegraphics[width=\textwidth]{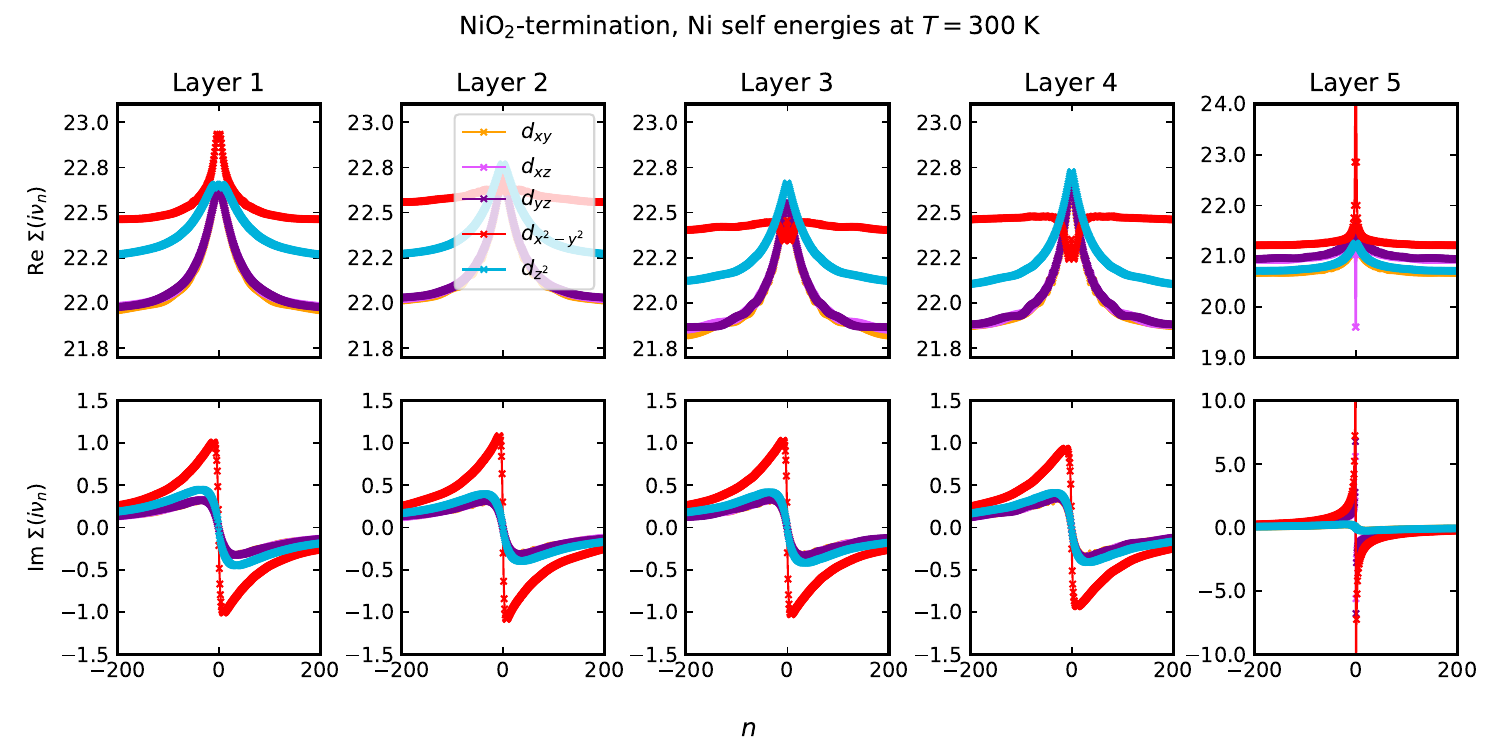}
     \end{subfigure}
     \hfill
     \begin{subfigure}[b]{\textwidth}
         \centering
         \includegraphics[width=\textwidth]{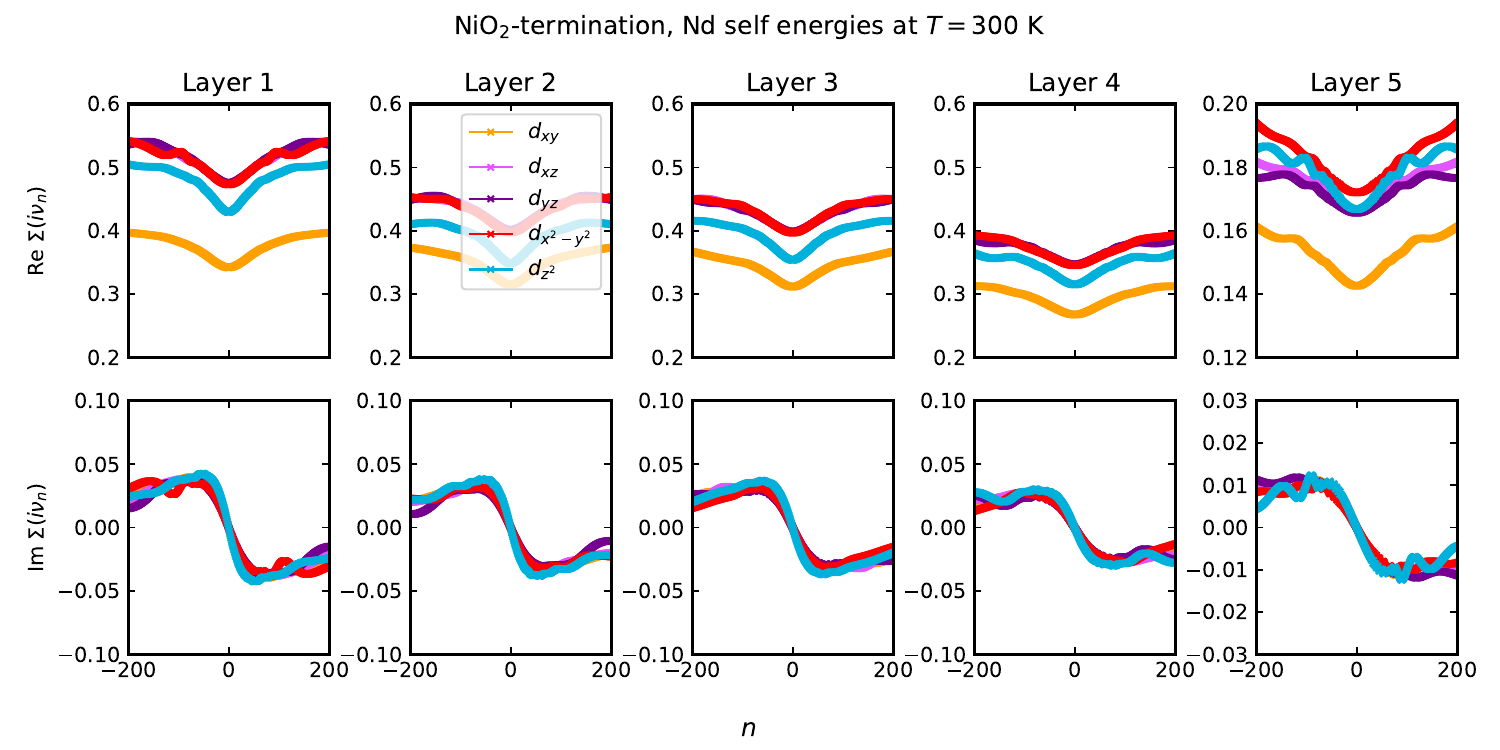}
     \end{subfigure}
        \caption{Layer-resolved self energy $\Sigma(i\nu_n)$ (real and imaginary part) for the NiO$_2$ termination (same structure as in main text) for Ni $3d$ (upper) and Nd $5d$ (lower) orbitals at $T=300$\,K, with the $n$-th fermionic Matsubara frequency $\nu_n = (2n+1)\pi \, T$. Mind the larger  scales of the plots in the surface region (layer 5), and  the much smaller scale of the Nd self energy.}
        \label{fig:siw_Nio2}
    \end{figure}

For the better reproducibility of our results, the layer-resolved self energy $\Sigma(i\nu_n)$ at $T=300$\,K, calculated with \textsc{w2dynamics}, is depicted in Fig.~\ref{fig:siw_Nio2}. The self energy of the  Ni $d$ orbitals located at the surface (layer 5) already suggests an insulating behavior of the surface $d_{x^2-y^2}$ and $d_{xz/yz}$ orbitals.
Please note that continuous-time QMC calculations tend to have an increasing error in the self energy 
for large frequencies. Hence, the large frequency values shown in  Fig.~\ref{fig:siw_Nio2} are to be taken with a grain of salt.

\begin{figure} [tb]
        \centering
        \includegraphics[width=\linewidth]{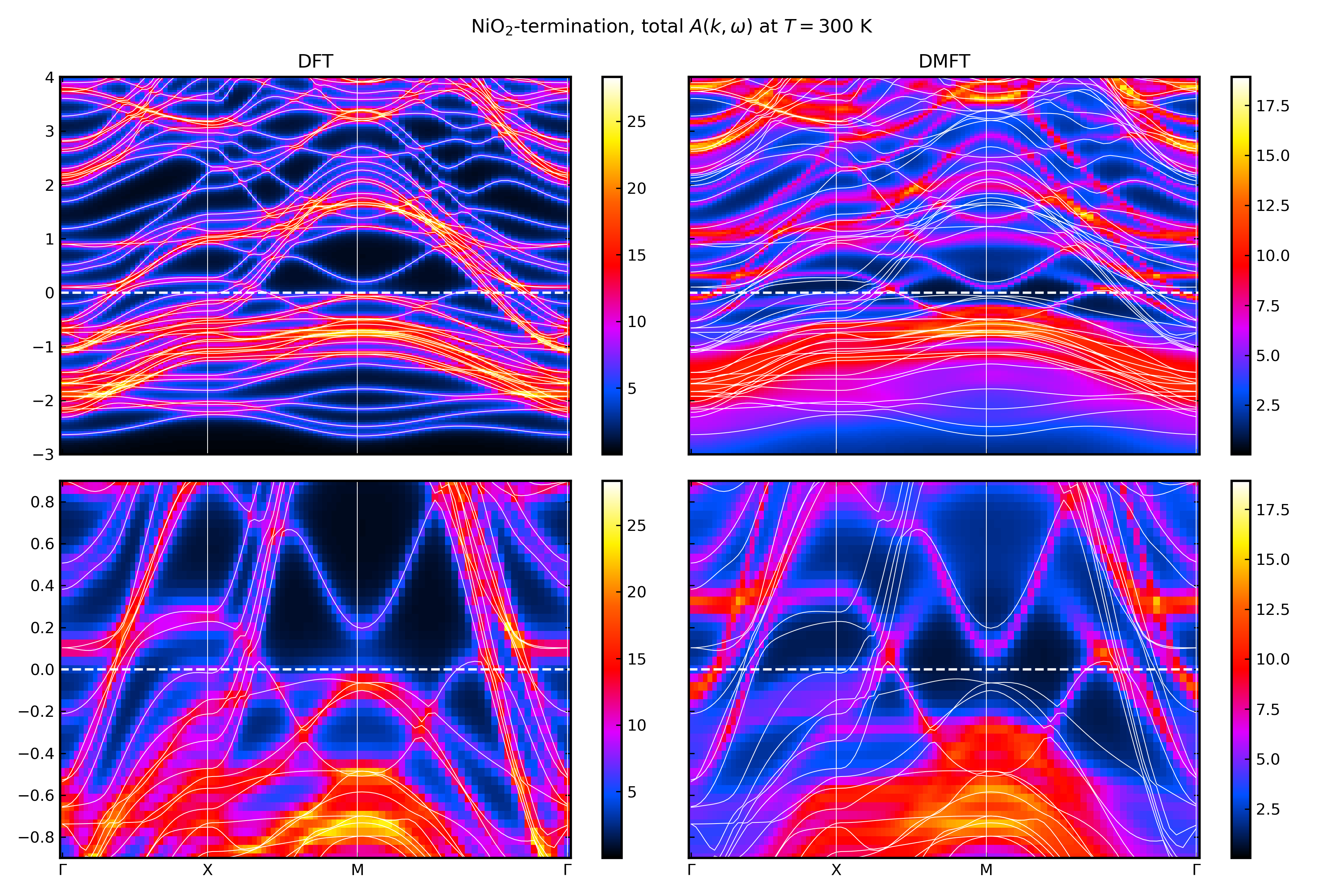}
        \caption{\label{fig:Awk_NiO2}Total spectral function of the NiO$_2$-terminated surface (same structure as in main text) obtained by DFT (left) and DMFT (right). White lines are corresponding DFT bands of the non-interacting Hamiltonian. Lower panels are the same as upper panels but magnified on a smaller energy range.}
    \end{figure}

\begin{figure} [tb]
        \centering
        \includegraphics[width=\linewidth]{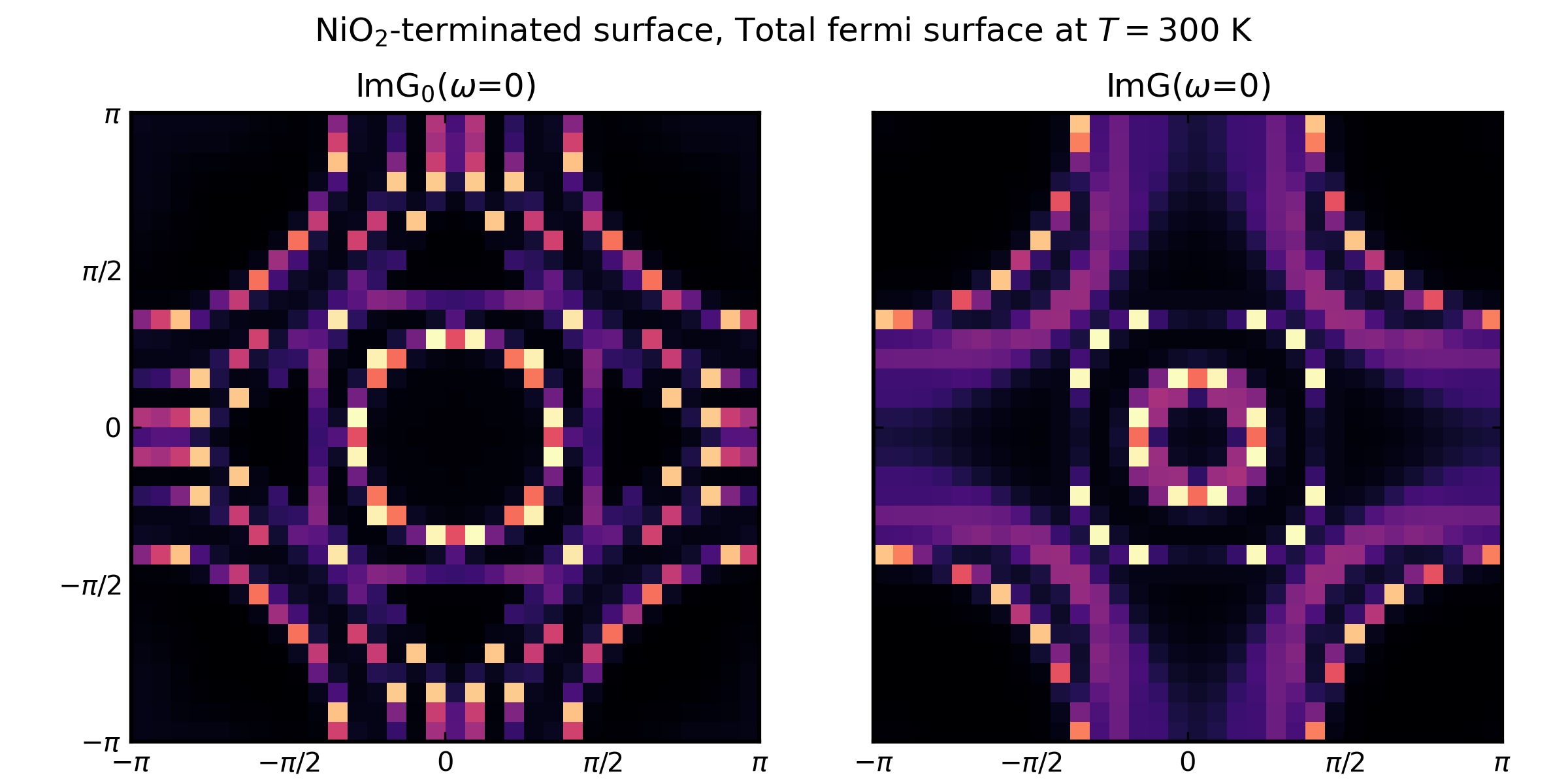}
        \caption{\label{fig:total_fs_NiO2}Total 2D Fermi surface of NiO$_2$ terminated surface (same structure as in main text) obtained by DFT  with Lorentzian broadening (left) and DMFT (right).}
    \end{figure}

In addition to the Nd and Ni contributions shown in the main text, Fig.~\ref{fig:Awk_NiO2}
shows the total ${\mathbf k}$-resolved DMFT spectral function 
including all layers and orbitals  and compares it to  DFT.
Fig.~\ref{fig:total_fs_NiO2} displays the corresponding total Fermi surface next to that obtained from the DFT band structure, or more precisely the DFT-derived Wannier Hamiltonian  with a Lorentzian broadening of 0.1\,eV.

\begin{figure} [tb]
        \centering
        \includegraphics[width=\linewidth]{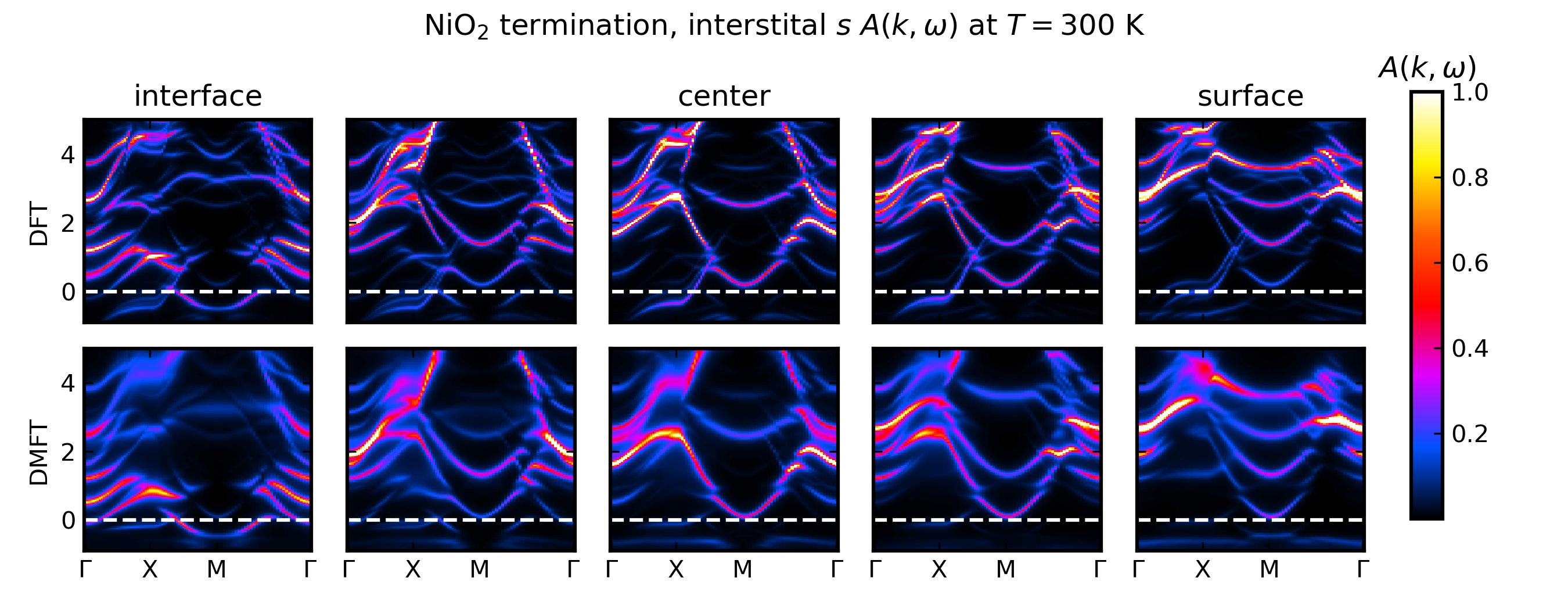}
        \caption{\label{fig:Awk_s_NiO2}Layer-resolved, $\mathbf{k}$-dependent spectral function of interstitial $s$ orbitals in the NiO$_2$-terminated surface (same structure as in main text) for DFT (upper row) and DMFT (bottom row). Left column is at the interface, right column is at the surface.}
    \end{figure}

Fig.~\ref{fig:Awk_s_NiO2} shows  the spectral function of the interstitial $s$ orbital that was not presented in the main text. Since the interstitial $s$ orbitals are treated as non-correlated, their change is only secondary due to changes in hybridization with the correlated Ni and Nd orbitals.

\begin{figure} [tb]
        \centering
        \includegraphics[width=\linewidth]{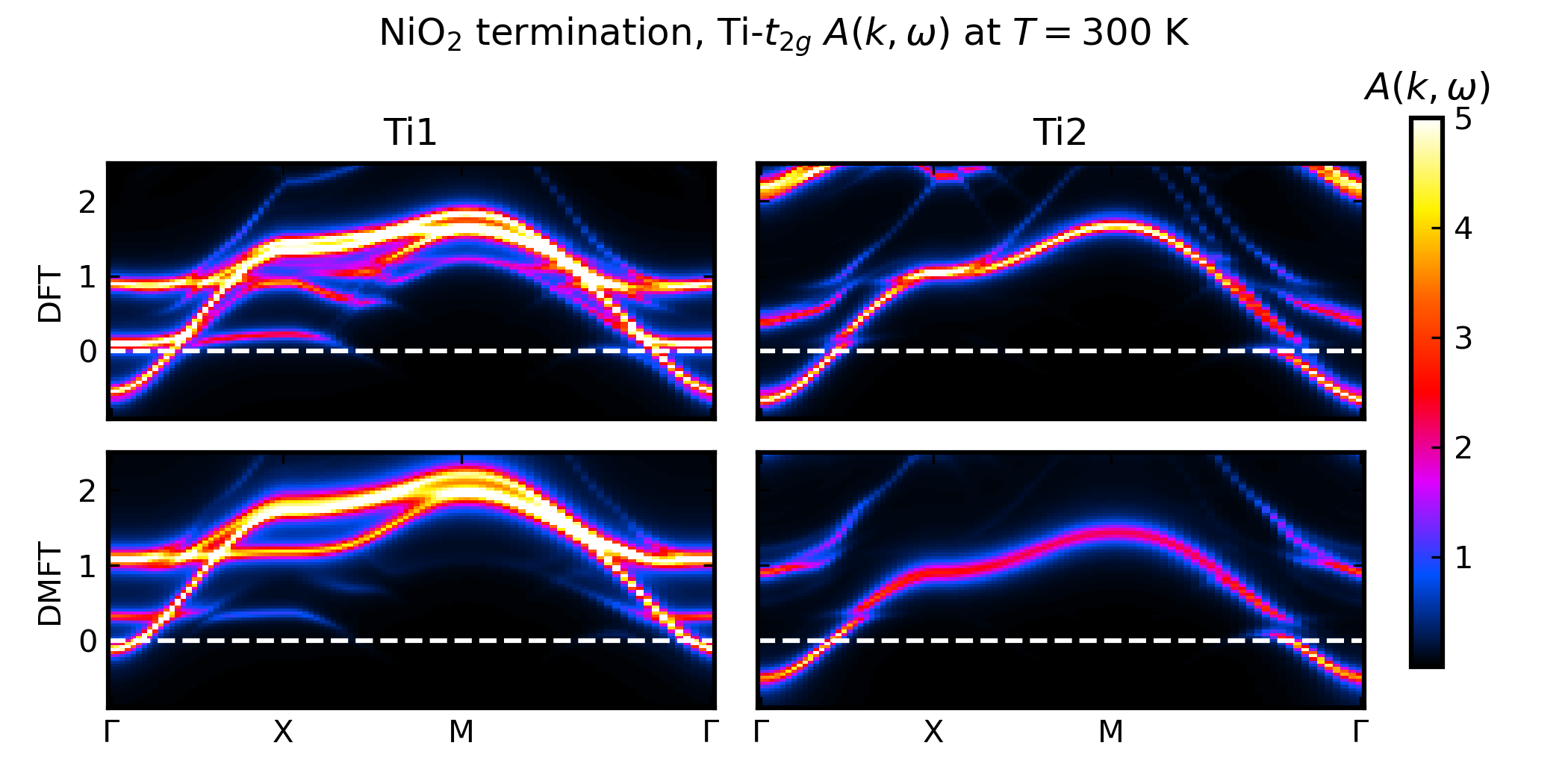}
        \caption{\label{fig:Awk_Ti_NiO2}Layer-resolved, $\mathbf{k}$-dependent spectral function of the Ti $t_{2g}$ manifold for the NiO$_2$-terminated surface (same structure as in main text) in DFT (upper row) and DMFT (bottom row). Left column is lowest Ti, right column is Ti in contact with NdNiO$_2$ film.}
    \end{figure}

Fig.~\ref{fig:Awk_Ti_NiO2} further  presents the Ti spectral functions of the two substrate layers included in the slab. The Ti $t_{2g}$ bands cross the Fermi energy and are partially occupied in DFT. For the NiO$_2$-surface termination, there are 2.35 electrons in both Ti layers combined. Including local correlations on the level of DMFT shifts them slightly upward, resulting in a reduced filling of 1.76 electrons. However, as Fig.~\ref{fig:Awk_Ti_NiO2} reveals, they still cross the Fermi energy and are $n$-type doped.

\clearpage 

\subsection{Nd-terminated slab}
\label{sec:add_figs_nd}

\begin{figure}[tb]
     \centering
     \begin{subfigure}[b]{\textwidth}
         \centering
         \includegraphics[width=\textwidth]{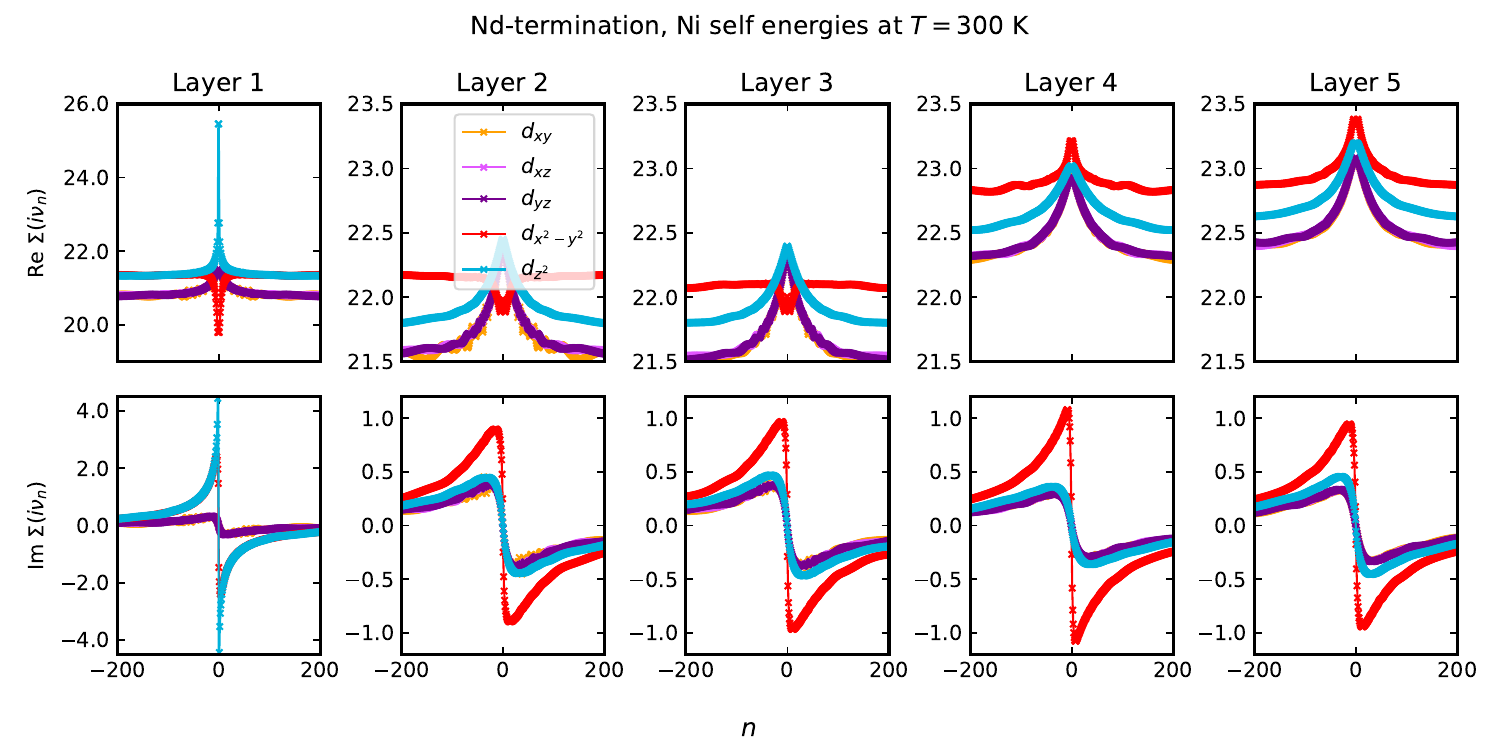}
     \end{subfigure}
     \hfill
     \begin{subfigure}[b]{\textwidth}
         \centering
         \includegraphics[width=\textwidth]{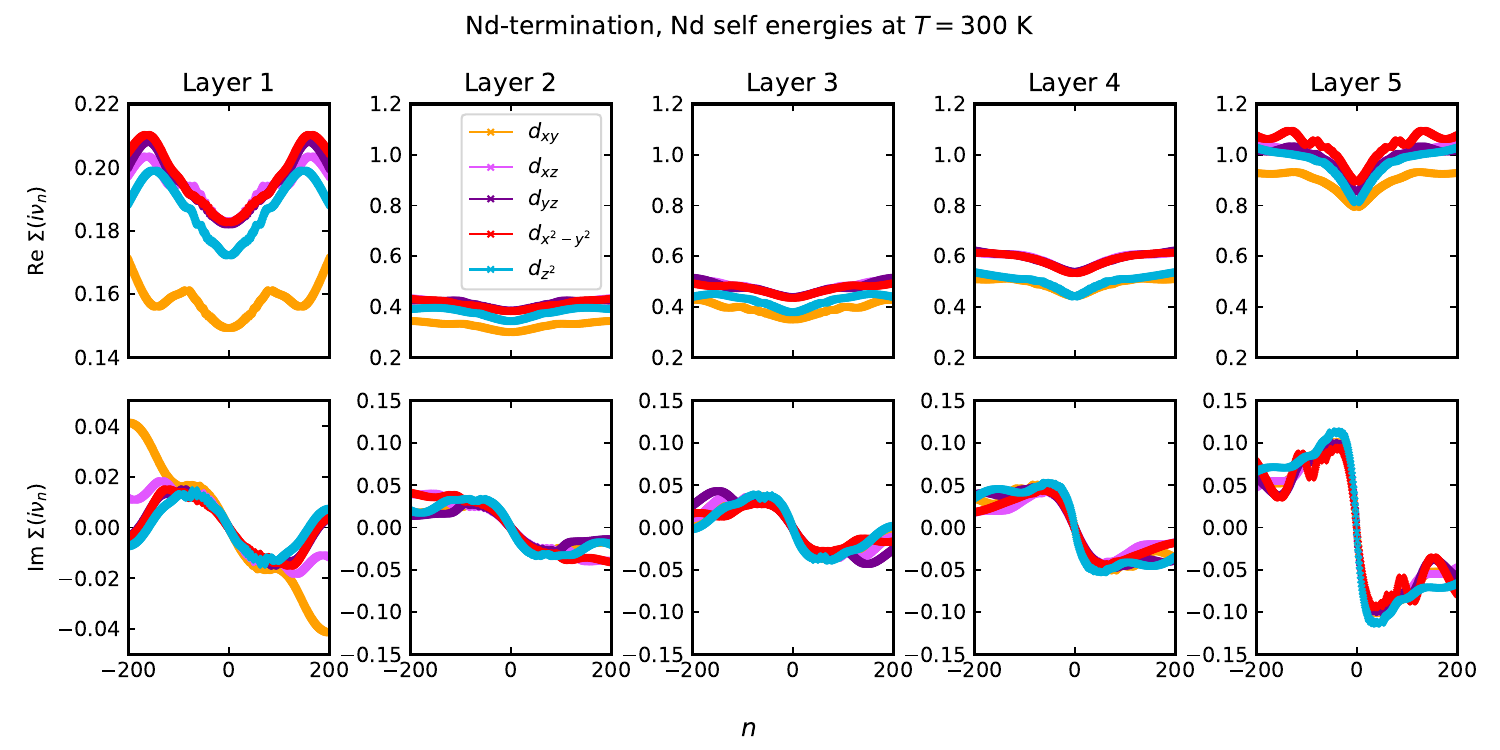}
     \end{subfigure}
        \caption{Same as Fig.~\ref{fig:siw_Nio2} but for the Nd-terminated slab in the main text. Mind again the different scales of the plots in the interface region (layer 1) and for the Nd self energy.}
        \label{fig:siw_Nd}
    \end{figure}

For the slab with Nd-surface termination, also presented in the main text, the NiO$_2$ interface (layer 1) becomes an orbital selective insulator with the formation of two Hubbard bands in the Ni and Ni $3d_{z^2}$ orbital. This can again be seen from the imaginary frequency self energy at the interface in Fig.~\ref{fig:siw_Nd} (upper left panels).

\begin{figure} [tb]
        \centering
        \includegraphics[width=\linewidth]{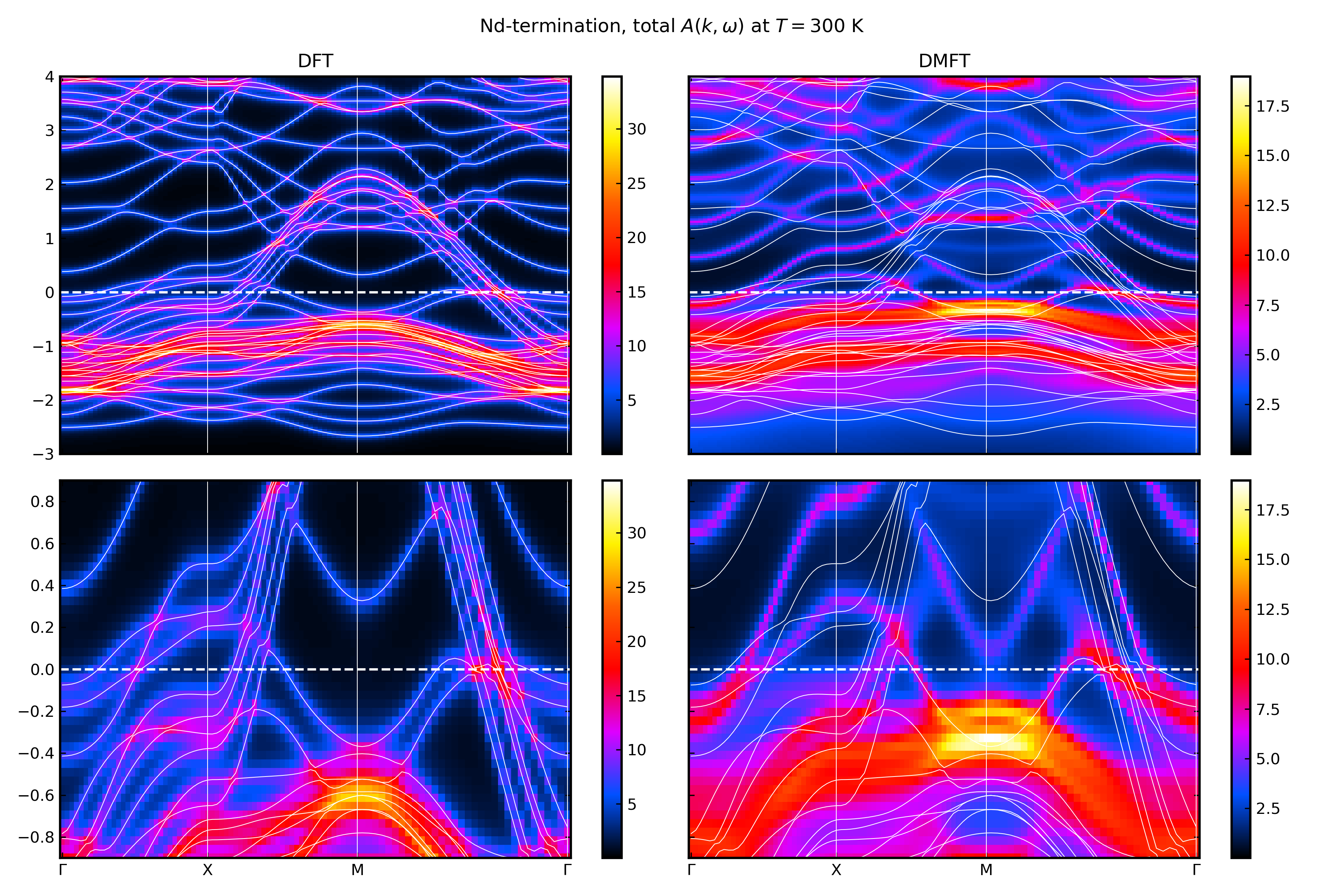}
        \caption{\label{fig:Awk_Nd}Same as Fig.~\ref{fig:Awk_NiO2} but for the Nd-terminated slab in the main text.}
    \end{figure}

\begin{figure} [tb]
        \centering
        \includegraphics[width=\linewidth]{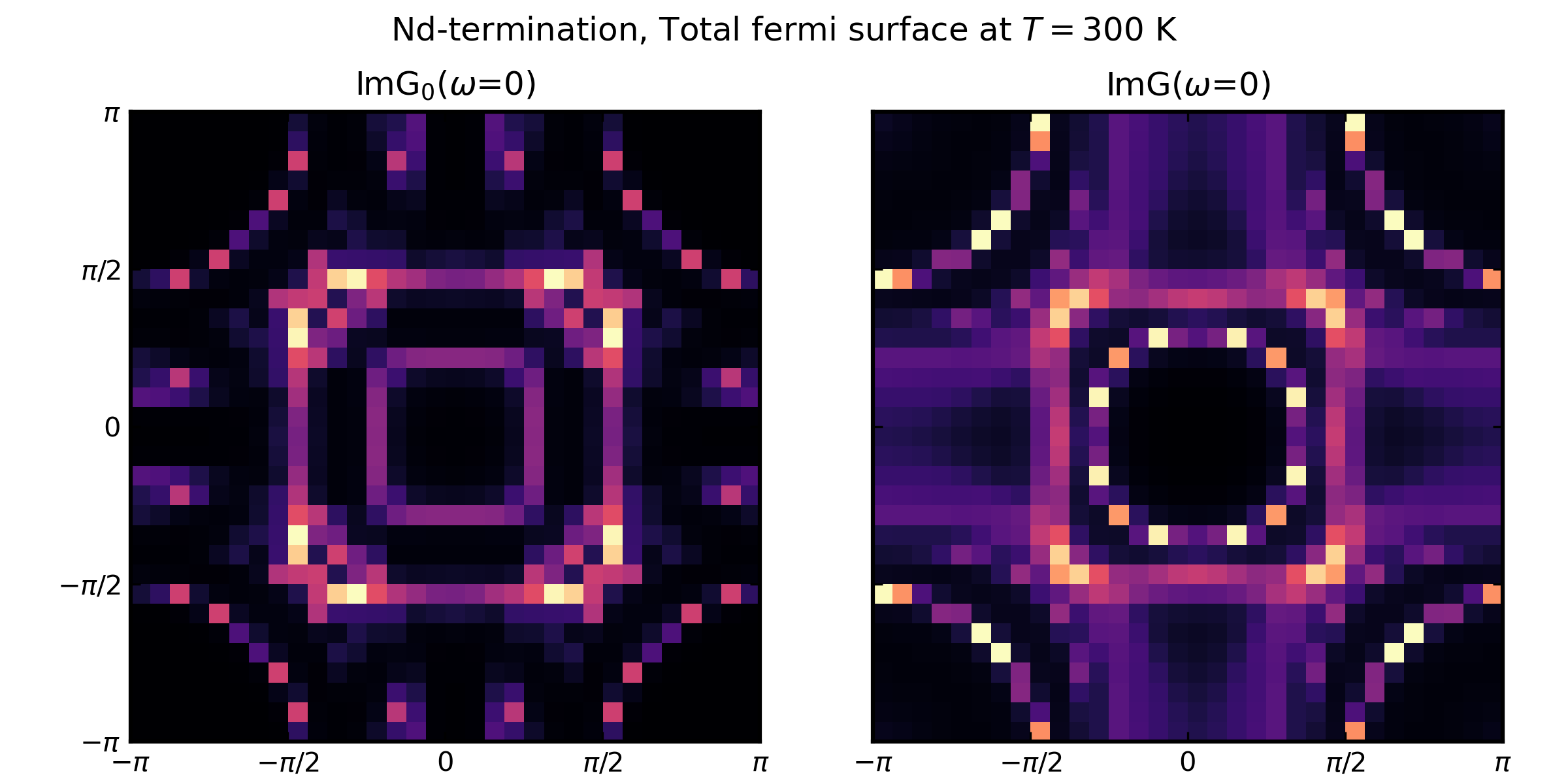}
        \caption{\label{fig:total_fs_Nd}Same as Fig.~\ref{fig:total_fs_NiO2} but for the Nd-terminated slab in the main text.}
    \end{figure}

\begin{figure} [tb]
        \centering
        \includegraphics[width=\linewidth]{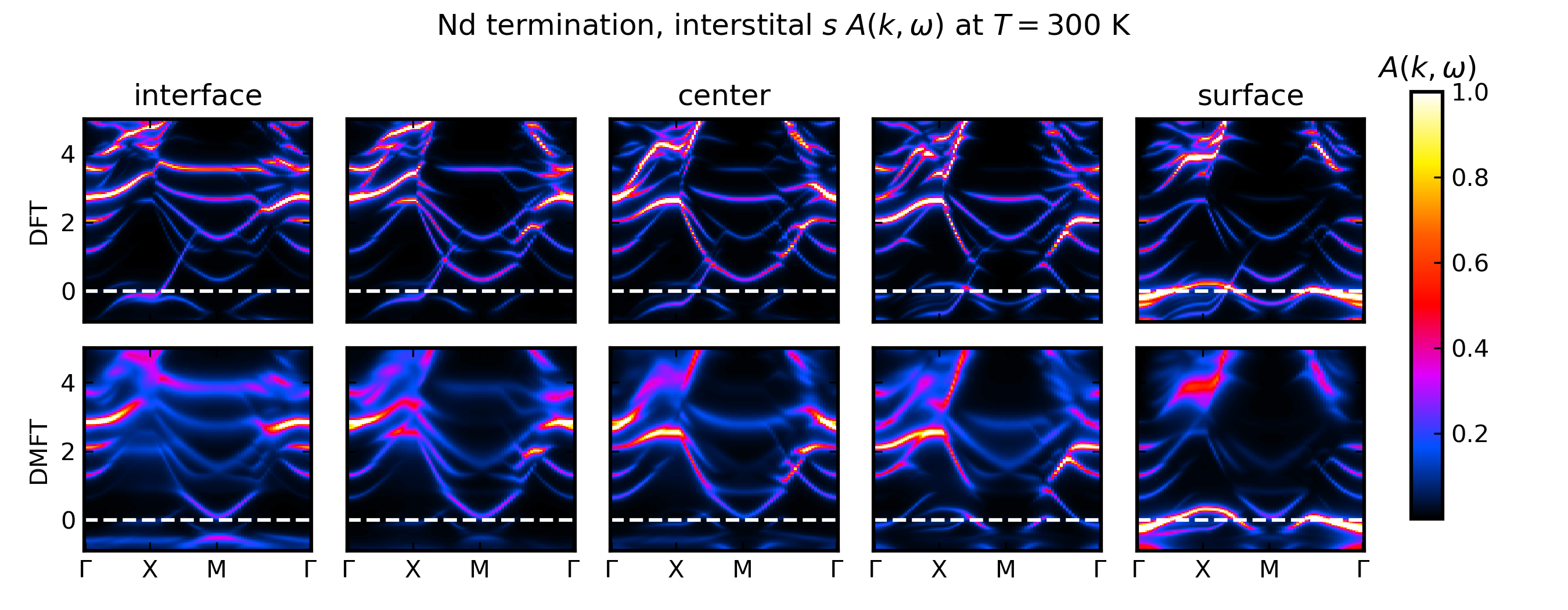}
        \caption{\label{fig:Awk_s_Nd}Same as Fig.~\ref{fig:Awk_s_NiO2} but for the Nd-terminated slab of the main text.}
    \end{figure}

    The total ${\mathbf k}$-resolved spectrum is shown in  Fig.~\ref{fig:Awk_Nd} for the Nd-surface termination and the corresponding Fermi surface in Fig.~\ref{fig:total_fs_Nd}.
    The interstitial $s$ spectral function is shown layer-resolved in Fig.~\ref{fig:Awk_s_Nd}.
    For the Nd termination, no Ti bands cross the Fermi energy. Hence, there was no need to take them into account in the DMFT calculations and there is no figure with their spectral function.
    

    \clearpage
\section{Other slabs}
\label{sec:other_slabs}

In this section, we present further supplemental calculations testing various parameters, including a larger slab, a freestanding slab (without SrTiO$_3$ substrate), different temperatures, and a simplified slab Hamiltonian, created from the bulk hoppings.

\subsection{Larger slab (7 layers)}
\label{sec:7layers}
\begin{figure}[tb]
     \centering
     \begin{subfigure}[b]{0.49\textwidth}
         \centering
         \includegraphics[width=\textwidth]{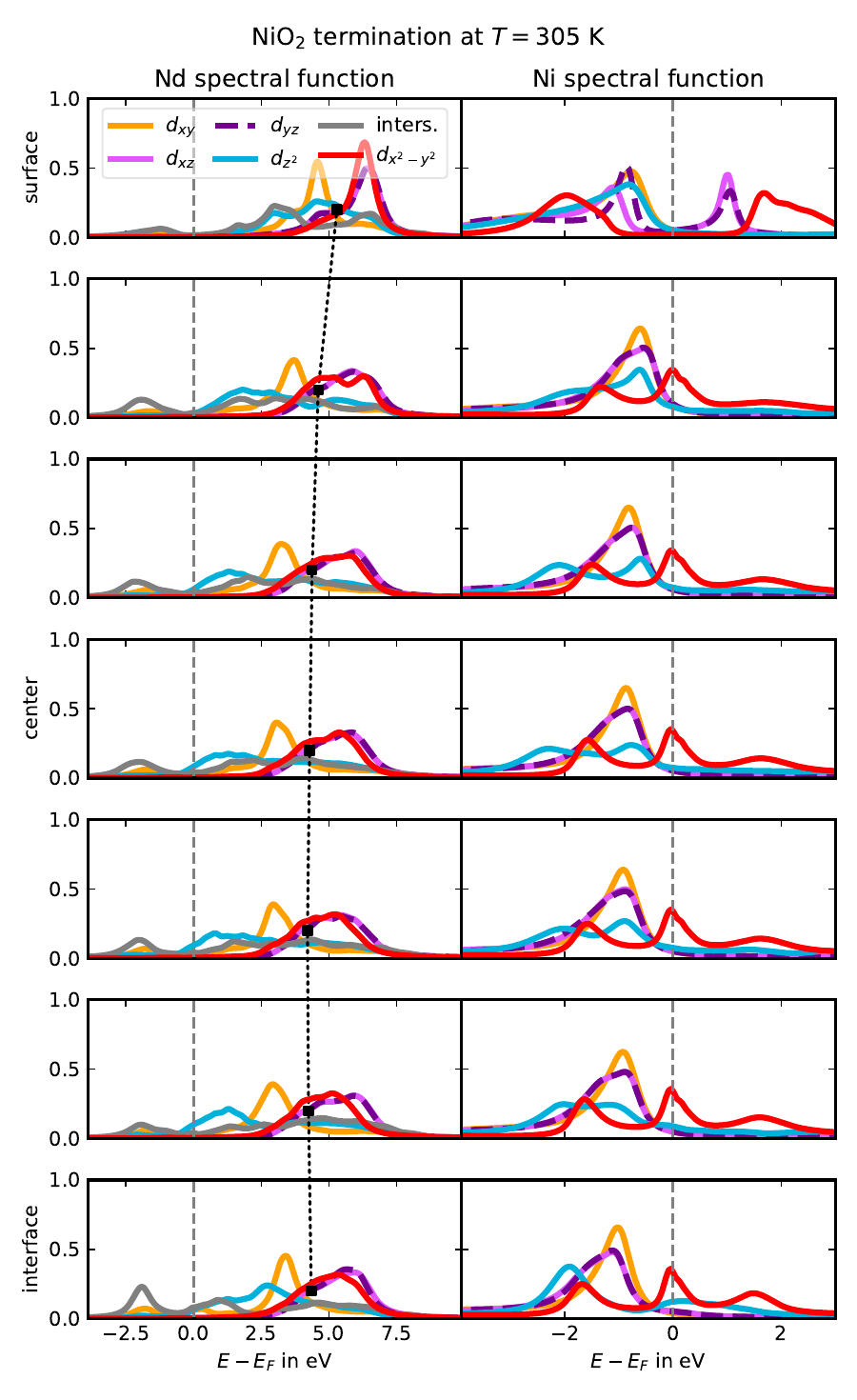}
     \end{subfigure}
     \hfill
     \begin{subfigure}[b]{0.49\textwidth}
         \centering
         \includegraphics[width=\textwidth]{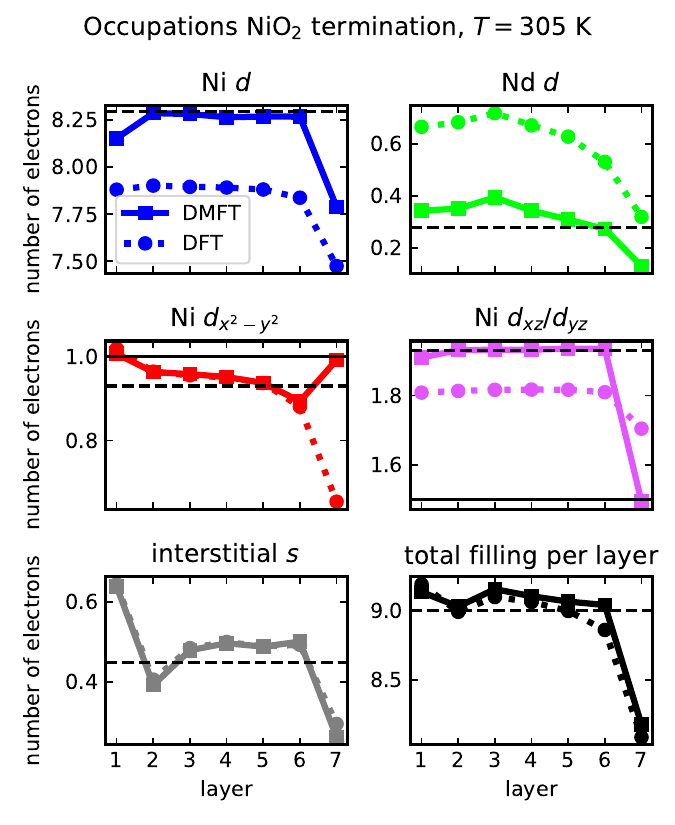}
     \end{subfigure}
         \centering
    \caption{\label{fig:7_layers_NiO2}Slab consisting of 7 NdNiO$_2$ layers and 2 SrTiO$_3$ layers, at $T=305$\,K. Left: Layer-resolved, $\mathbf{k}$-integrated spectral function. Right: Filling in DMFT (solid lines, squares) and DFT (dotted lines, circles). Dashed, black horizontals are the fillings from bulk DMFT calculations of NdNiO$_2$. Black, solid lines in the central row indicate half-filling of Ni $3d_{x^2-y^2}$ and three-quarter filling of the degenerate Ni $3d_{xy}/d_{yz}$ orbitals, respectively.}
    \end{figure}

\begin{figure}[tb]
     \centering
     \begin{subfigure}[b]{\textwidth}
         \centering
         \includegraphics[width=\textwidth]{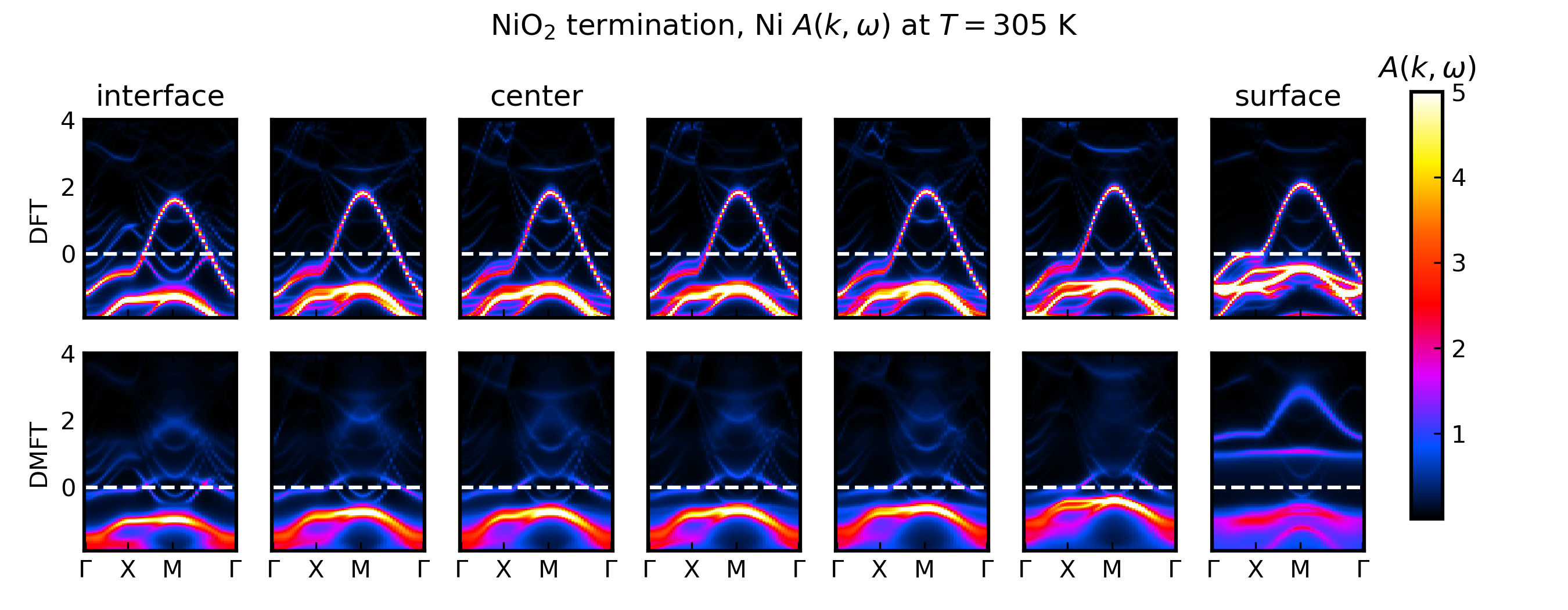}
     \end{subfigure}
     \begin{subfigure}[b]{1\textwidth}
         \centering
         \includegraphics[width=\textwidth]{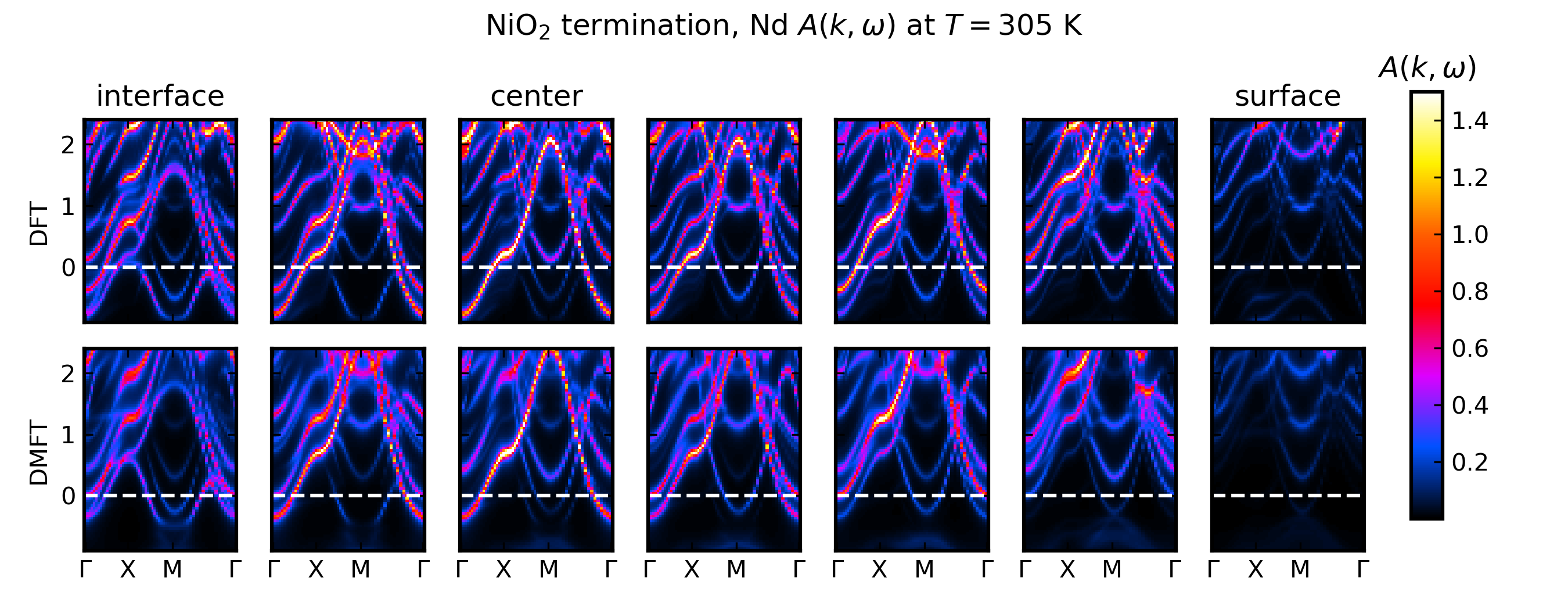}
     \end{subfigure}
     \begin{subfigure}[b]{1\textwidth}
         \centering
         \includegraphics[width=\textwidth]{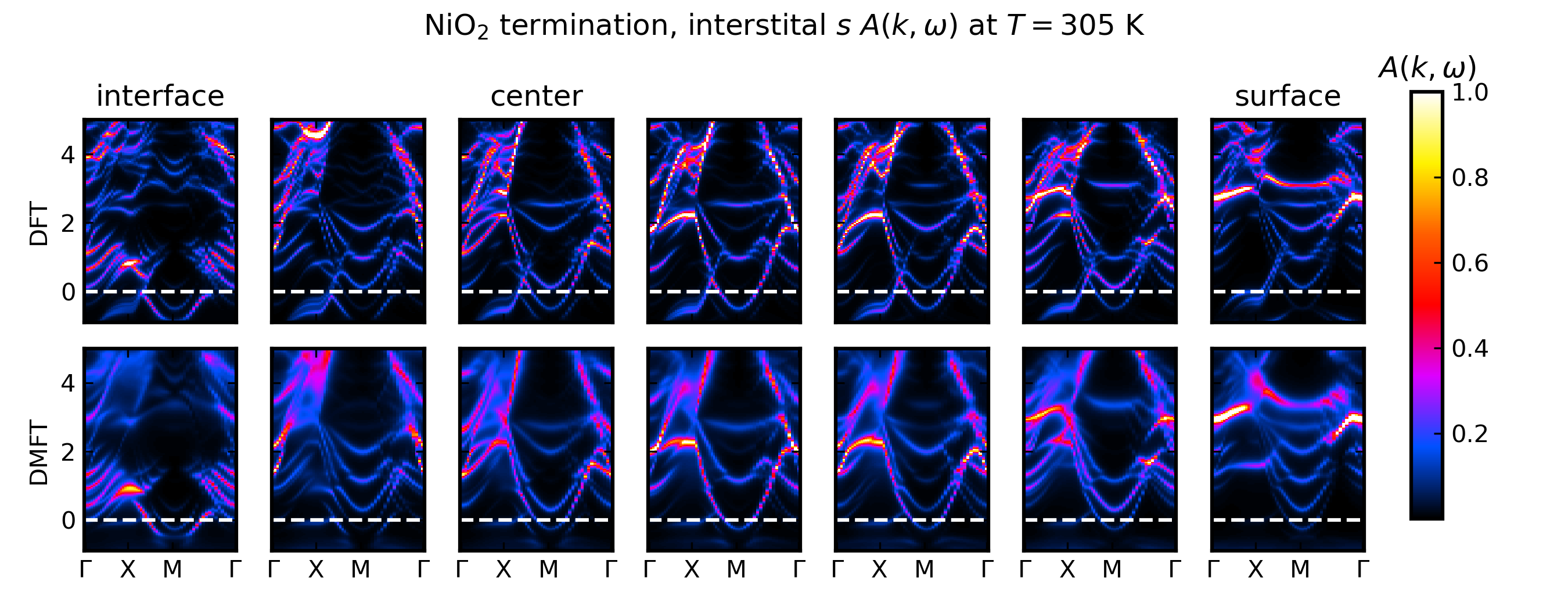}
     \end{subfigure}
        \caption{Same as Fig.~\ref{fig:Awk_s_Nd} for NiO$_2$-terminated slab but with 7 NdNiO$_2$ layers, projected onto Ni $3d$ (top), Nd $5d$ (center) and interstitial $s$ (bottom) orbitals at $T=305$\,K.}
        \label{fig:Awk_7_NiO2}
    \end{figure}

\begin{figure}[tb]
        \centering
        \includegraphics[width=\textwidth]{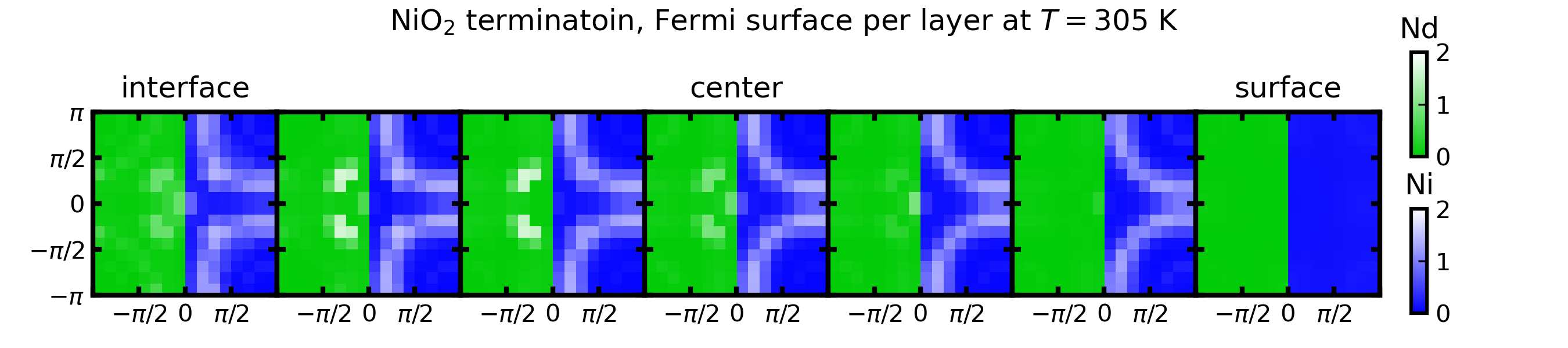}
        \caption{\label{fig:FS_layer_7_NiO2}Contribution of Nd (green) and Ni (blue) orbitals to the Fermi surface of NiO$_2$ terminated slab with 7 NdNiO$_2$ layers at $T=305$\,K.}
    \end{figure}

Here we consider the same setting as the NiO$_2$-terminated, 5 layered slab in the main text, but now with two additional layers of NdNiO$_2$. The calculation is carried out at $T=305$\,K for 97 orbitals (35 Ni $d$, 35 Nd $d$,  10 Sr $d$, 10 Ti $d$, 7 interstitial $s$).
Figs.~\ref{fig:7_layers_NiO2}-\ref{fig:FS_layer_7_NiO2} are the same kind of figures as presented in the main text but now with the aforementioned seven instead of five  NdNiO$_2$ layers. The physics is qualitatively very similar. Again, in 
Fig.~\ref{fig:7_layers_NiO2} (right), the trend of the orbital occupation changes dramatically for the surface layer with the Ni $d_{x^2-y^2}$ orbital becoming half-filled and a second hole in the two three-quarter filled Ni $d_{xz}$ and $d_{xz}$ orbitals.  As for five NdNiO$_2$ layers, this surface layer is insulating in Fig.~\ref{fig:7_layers_NiO2} (left) for seven layers, too.

Similarly to the five-layer case, in Fig.~\ref{fig:FS_layer_7_NiO2} there is no $\Gamma$ pocket for the surface and subsurface layer, whereas it is present at the interface layer.
We thus conclude that our main findings do not depend on the precise number of layers taken into account. As a matter of course it would be instructive to also have calculations for an even larger number of layers. However, the Wannierization turned out to be hard for even larger slabs, which is why we have restricted ourselves to five and seven layers.

    \clearpage
    \pagebreak
    
\subsection{Freestanding film (no substrate)}
\label{Sec:NoSTO}

\begin{figure}[tb]
     \centering
     \begin{subfigure}[b]{0.49\textwidth}
         \centering
         \includegraphics[width=\textwidth]{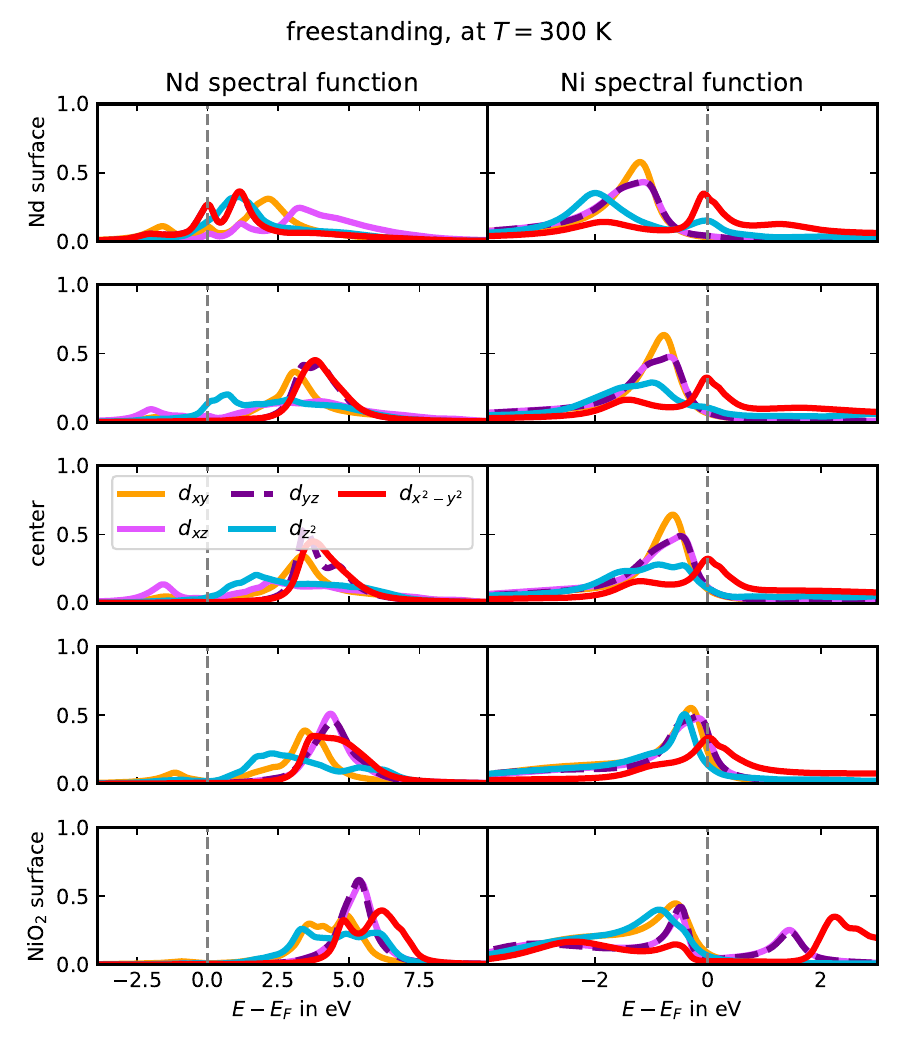}
     \end{subfigure}
     \hfill
     \begin{subfigure}[b]{0.49\textwidth}
         \centering
         \includegraphics[width=\textwidth]{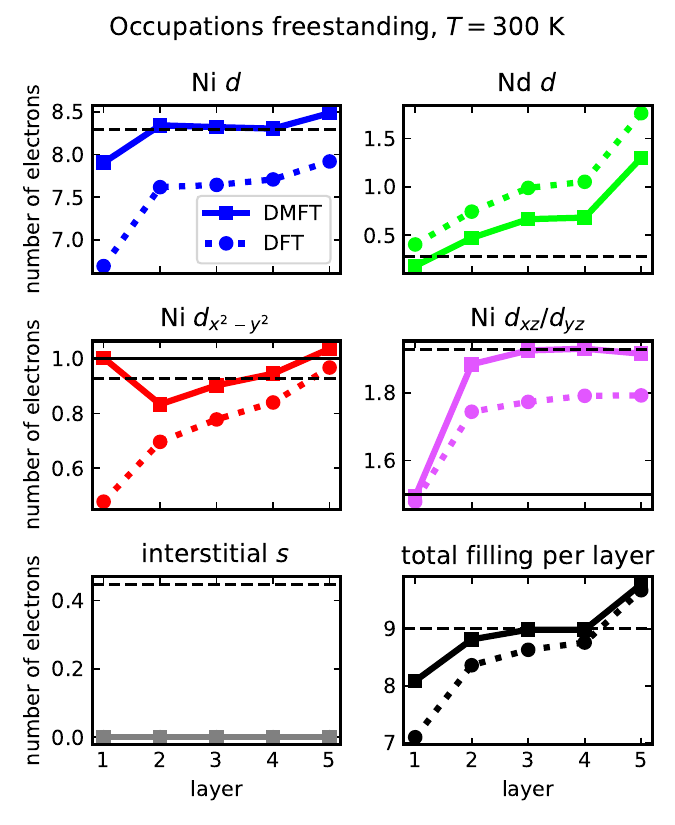}
     \end{subfigure}
        \caption{\label{fig:noSTO}Same as Fig.~\ref{fig:7_layers_NiO2}, but for 5 NdNiO$_2$ layers without SrTiO$_3$ substrate at $T=300$\,K. Layer 1 is the NiO$_2$-surface, Layer 5 is the Nd-surface.}
\end{figure}

\begin{figure} [tb]
        \centering
        \includegraphics[width=1\linewidth]{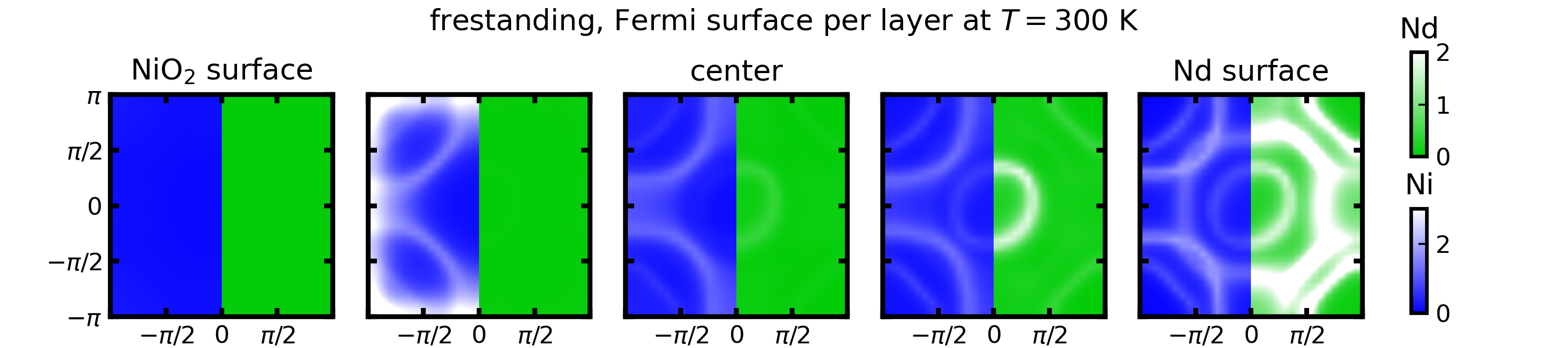}
        \caption{\label{fig:FS_layer_noSTO}Same as Fig.~\ref{fig:FS_layer_7_NiO2}, but for 5 NdNiO$_2$ layers without SrTiO$_3$ substrate at $T=300$\,K. The NiO$_2$-surface is to the left and the Nd-surface to the right.}
    \end{figure}

The results shown in Figs.~\ref{fig:noSTO} and \ref{fig:FS_layer_noSTO} are obtained for a freestanding, 5 layer NdNiO$_2$ film at $T=300$\,K. That is as in Fig.~1 of the main text (left or right)  but without the substrate layer in the copper-rose-shaded box.
After structural relaxation, the film conveniently exhibits both surface terminations in the same supercell (NiO$_2$-surface and Nd-surface). 

Fig.~\ref{fig:noSTO} (left) shows that the NiO$_2$ surface (which is akin to the surface layer for the NiO$_2$ termination and to the interface layer of the Nd termination) is insulating, whereas the Nd surface is metallic, as in the calculation with substrate and Nd-terminated surface. 
Given the square planar crystal field, the two holes are in the Ni $d_{x^2-y^2}$ and $d_{xz/yz}$ 
orbitals.

As before there is no $\Gamma$ pocket for the NiO$_2$ surface in Fig.~\ref{fig:FS_layer_noSTO}. However, there is a $\Gamma$ pocket and a more complex Fermi surface for the Nd surface.


\subsection{Different temperatures}
\label{sec:different_T}
For analyzing the prospective temperature dependence of our DMFT results, we study a considerably lower temperature $T=145$\,K, i.e., essentially  half of the room temperature that is presented in the main text. For the NiO$_2$-terminated slab, we also present results for $T=96\,$K.

\begin{figure}[tb]
     \centering
     \begin{subfigure}[b]{0.49\textwidth}
         \centering
         \includegraphics[width=\textwidth]{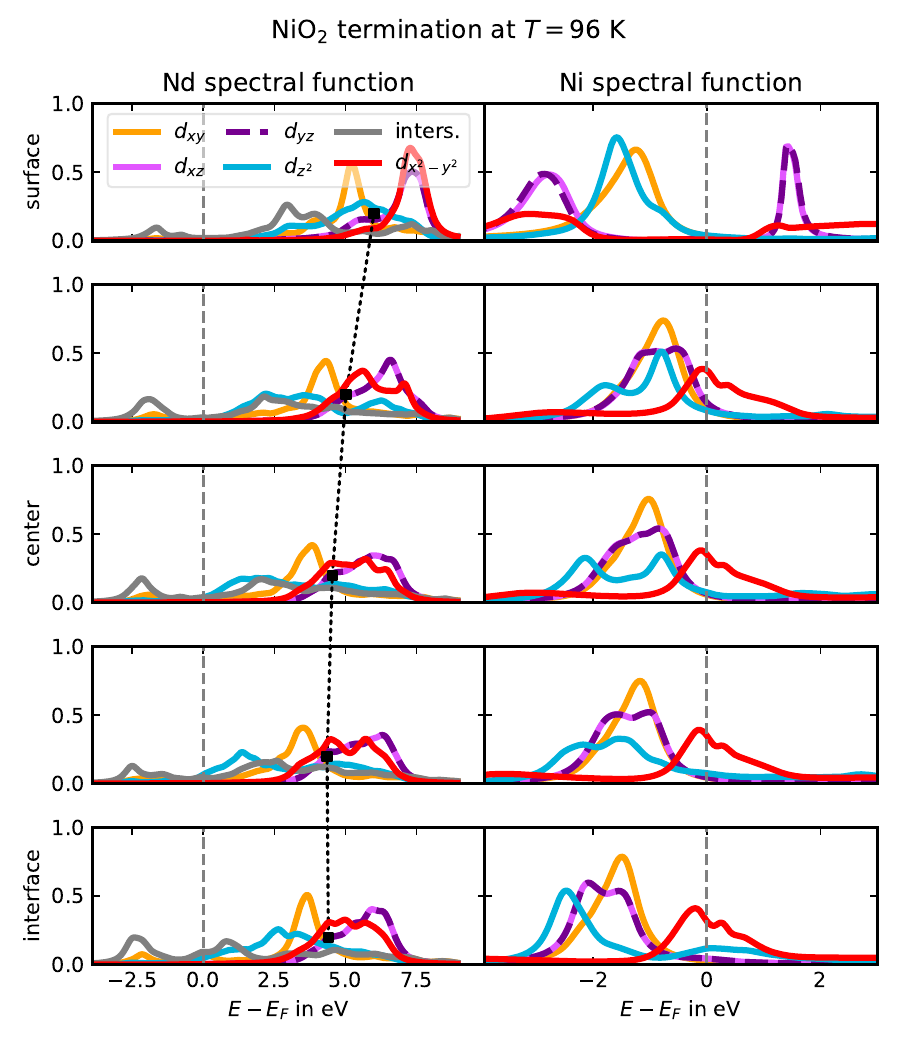}
     \end{subfigure}
     \hfill
     \begin{subfigure}[b]{0.49\textwidth}
         \centering
         \includegraphics[width=\textwidth]{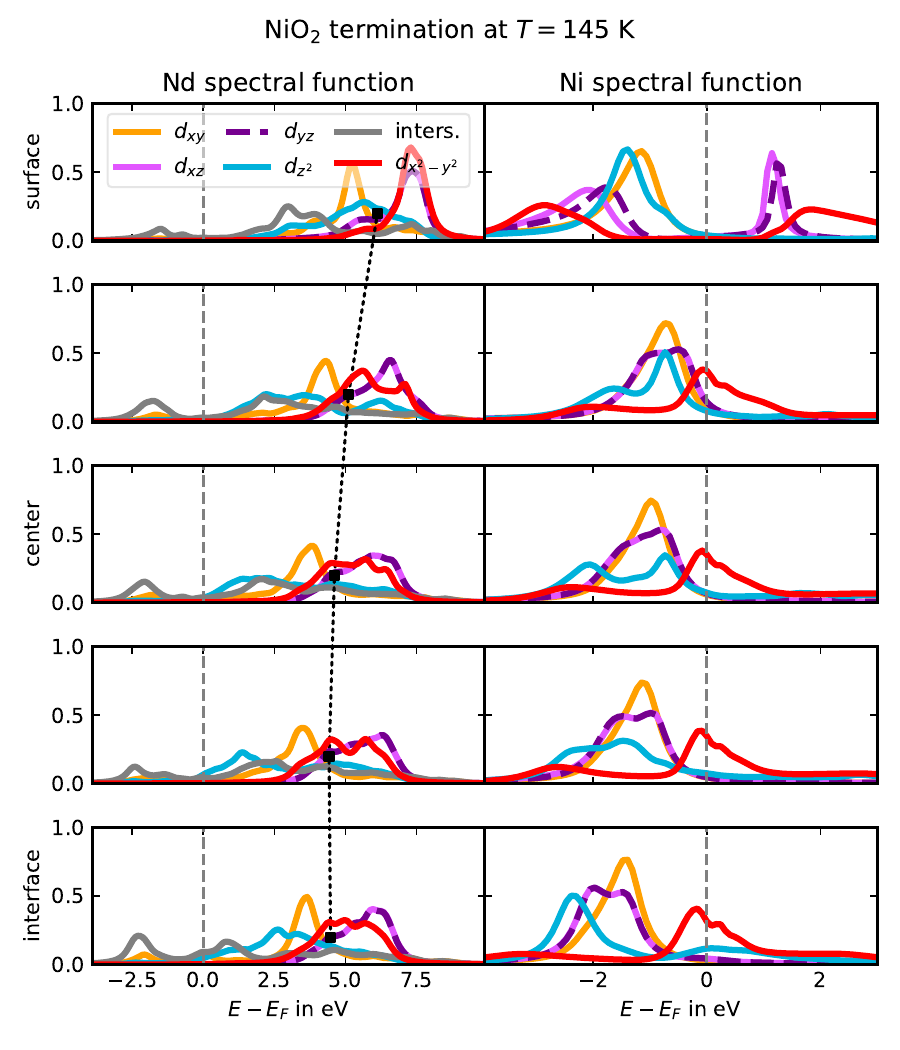}
     \end{subfigure}
        \caption{\label{fig:Aw_145K_NiO2}Layer-resolved spectral function of NiO$_2$-terminated slab from main text at $T=96$\,K (left) and $T=145$\,K (right). (Compare to Fig.~4 from main text.) }
\end{figure}

Fig.~\ref{fig:Aw_145K_NiO2} shows the NiO$_2$-terminated slab at $T=96$ and $145\,$K instead of 300\,K in Fig.~4 of the main text. The spectrum is very similar, illustrating that there is no major temperature dependence---except for the expected longer life times for a Fermi liquid at lower temperatures.

\begin{figure}[tb]
     \centering
     \begin{subfigure}[b]{0.49\textwidth}
         \centering
         \includegraphics[width=\textwidth]{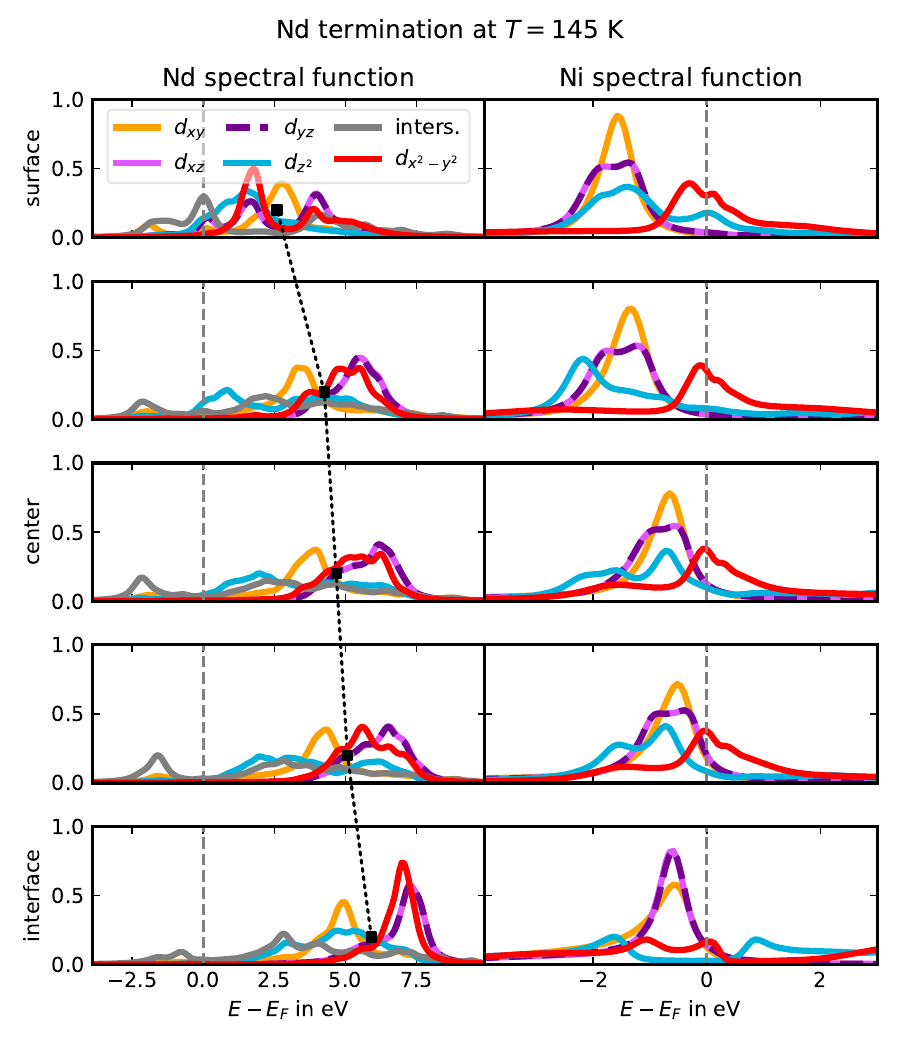}
     \end{subfigure}
     \caption{\label{fig:Aw_temperature_Nd}Layer-resolved spectral function of the Nd-terminated slab from main text but at $T=145$\,K. (Compare to Fig.~9 from main text.)}
\end{figure}

Fig.~\ref{fig:Aw_temperature_Nd} presents an additional temperature for the Nd-terminated slab, to be compared to Fig.~9 of the main text. Again, the qualitative features are the same. The slightly different sharpness of different peaks at different temperatures might also be an artifact induced by the errors of the maximum entropy method.


\subsection{Tight-binding Model}
\label{sec:tight_binding}
\begin{figure}
        \centering
        \includegraphics[width=1\linewidth]{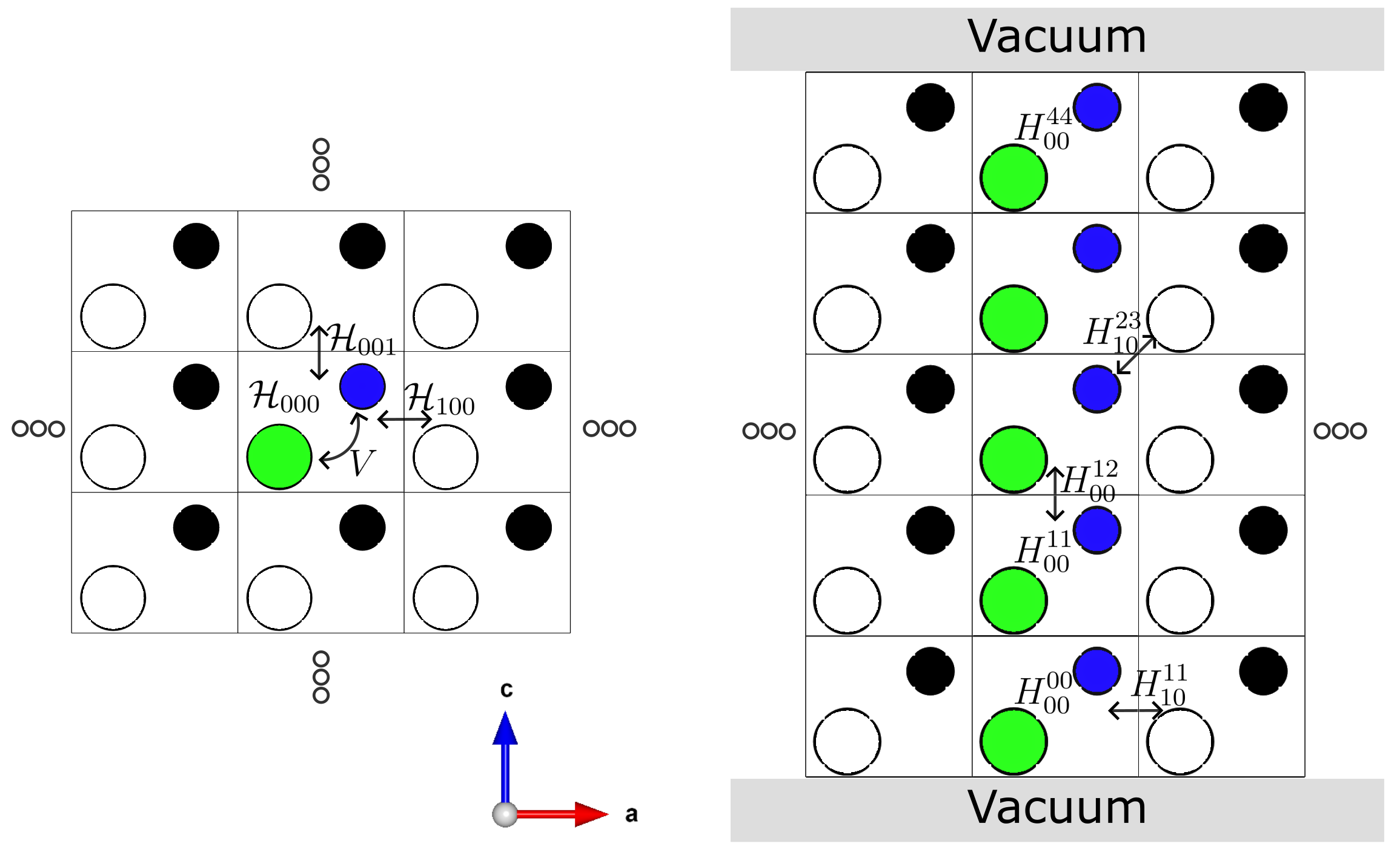}
        \caption{\label{fig:tb_hamiltonian} Sketch of bulk (left) and slab (right) real-space Hamiltonian components, as defined in Eqs.~\ref{eq:hamiltonian_bulk} and \ref{eq:hamiltonian_slab}. The first unit cell is indicated by green (Nd) and blue (Ni) atoms, while white and black atoms are from neighboring unit cells. Mind that $x$- and $y$-direction are equivalent. The oxygen atoms are not displayed as they are neglected in the tight binding Hamiltonian.}
\end{figure}

Further, we model the influence of vacuum on both ends of a pristine freestanding thin film as in Section~\ref{Sec:NoSTO}. However, instead of calculating the electronic structure of an explicit $1 \times 1 \times N$ slab system, here we construct an auxiliary $10N$-band slab Hamiltonian $H$ from the real space, 10 band bulk Hamiltonian $\mathcal{H}$:

\begin{align}
    \mathcal{H}_{000} =
    \left(
    \begin{array}{c|c}
       \varepsilon^{\text{Ni}} & V_{}  \\ \hline
         V^{\dagger}_{} & \varepsilon^{\text{Nd}} 
    \end{array} 
    \right) \in \mathbb{C}^{10 \times 10}
    \label{eq:hamiltonian_bulk}
\end{align}
where 
\begin{align}
    \varepsilon= \text{diag}( \varepsilon_{d_{xy}},  \varepsilon_{d_{xz}}, \varepsilon_{d_{yz}}, \varepsilon_{d_{x^2-y^2}}, \varepsilon_{d_{z^2}}) \in \mathbb{C}^{5 \times 5}
\end{align}
are onsite energies for Ni and Nd ion in the tetragonal unit cell and $V \in \mathbb{C}^{5 \times 5}$ are hybridizations between the sites. Inter-cell hoppings between all orbitals are covered in $\mathcal{H}_{lmn}$ with $lmn \neq 000$ (see Fig.~\ref{fig:tb_hamiltonian}). \\
The $1 \times 1 \times N$, effective 2D slab Hamiltonian is then constructed from the bulk Hamiltonian $\mathcal{H}$ as:

\begin{align}
    H_{lm}^{\alpha\beta} = \mathcal{H}_{lm(\beta-\alpha)} \in \mathbb{C}^{10 \times 10}.
    \label{eq:hamiltonian_slab}
\end{align}
The Greek indices $(\alpha,\beta=\{0,1,2,...,N-1\})$ denote the respective layer within the supercell, while $lm$ denote real-space coordinates in $x$- and $y$-direction, respectively. \\
This definition introduces an infinite vacuum along the $z$-axis, since it allows for hopping in $z$-direction only between layers within the supercell, making all layers non-equivalent. This model is hence a simplification of a freestanding film without substrate, where both (pristine and unrelaxed) surfaces are present in the same slab. The definitions of matrix elements are also summarized in Fig.~\ref{fig:tb_hamiltonian}.

Since the real film exhibits a polar field build up, we explicitly apply an electric field in $z$-direction, with potential $V$ between both surfaces. The electric field effectively reduces the onsite energy at the Nd-terminated surface and increases onsite energy at NiO$_2$-terminated surface.

\begin{figure}
    \centering
    \includegraphics[width=\linewidth]{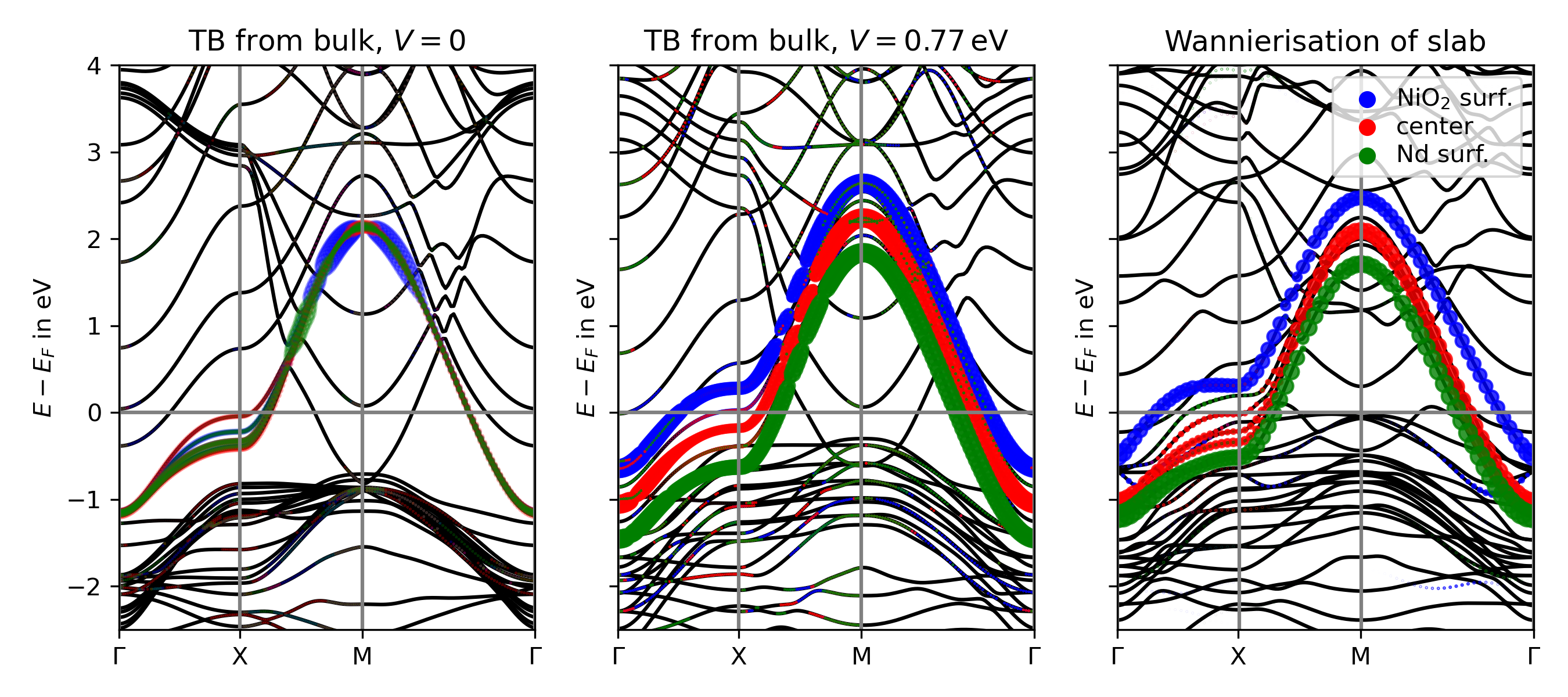}
    \caption{Band structure of $1 \times 1 \times 5$ slab constructed from the tight binding  10 band Hamiltonian without (left) and with an electric field between the two surfaces (center). Band structure of the full, DFT relaxed slab consisting of 5 layers and no substrate (right). Projections are onto the Ni$d_{x^2-y^2}$ orbital closest to the Nd-terminated surface (green), in the center (red) and at the NiO$_2$ surface.}
    \label{fig:TB_bands}
\end{figure}

For the $1 \times 1 \times 5$ Hamiltonian, the potential difference between the Nd- and NiO$_2$-terminated surface is heuristically set to $V = 0.77$\,eV, such that the splitting of the Ni-$d_{x^2-y^2}$ bands at the $M$ point roughly matches the splitting obtained for a DFT calculation of the corresponding 5 layer slab supercell (see Fig.~\ref{fig:TB_bands}).


\begin{figure}[tb]
     \centering
     \begin{subfigure}[b]{0.49\textwidth}
         \centering
        \includegraphics[width=\textwidth]{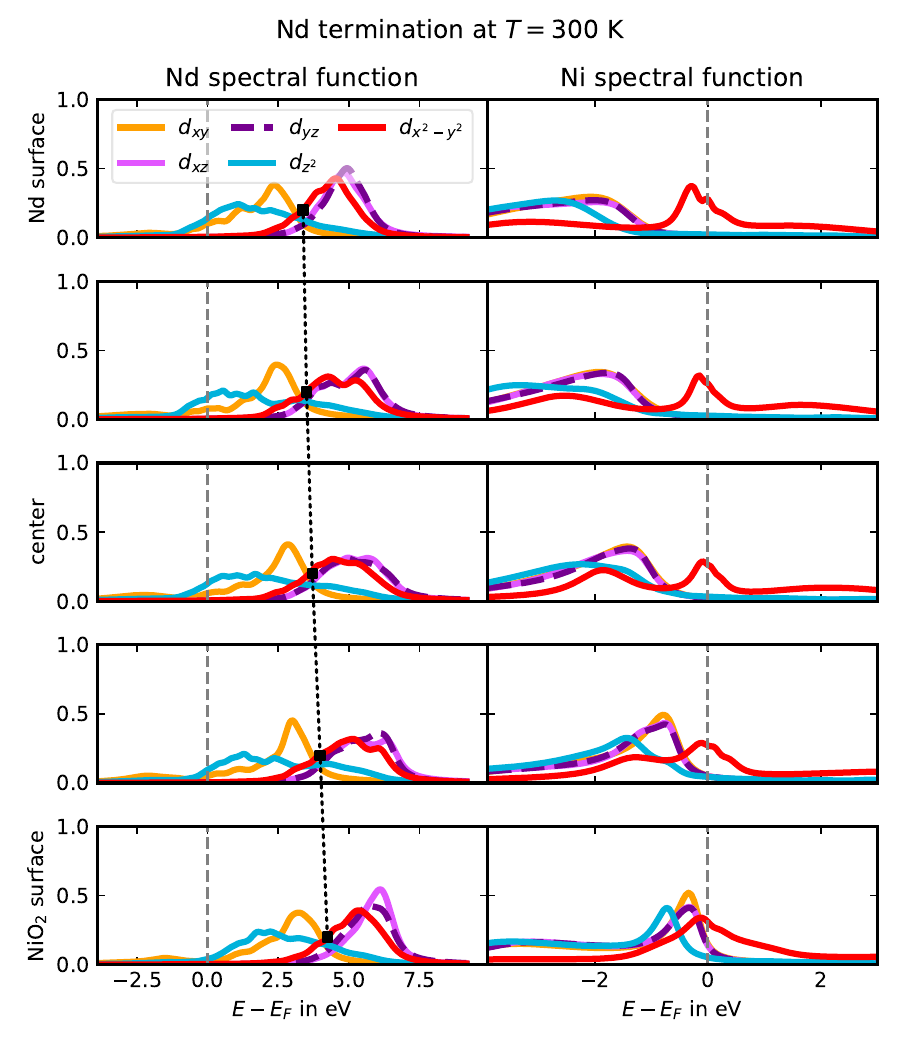}         
     \end{subfigure}
     \hfill
     \begin{subfigure}[b]{0.49\textwidth}
         \centering
         \includegraphics[width=\textwidth]{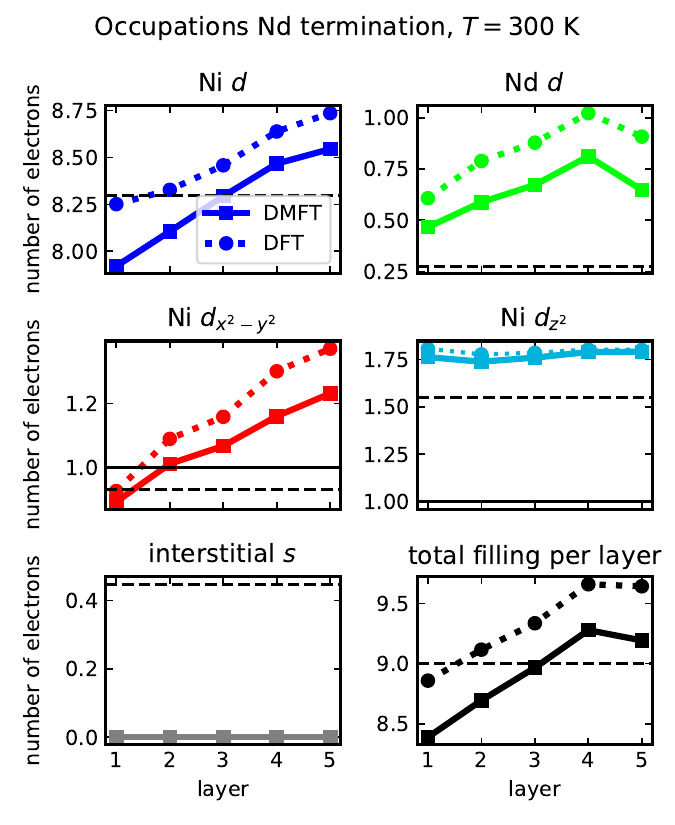}
     \end{subfigure}
        \caption{\label{fig:tb_dmft}Spectral function and filling for the tight-binding model with five NdNiO$_2$ layers that have a potential difference of $V=0.77$\,eV.}
\end{figure}

\begin{figure} [tb]
        \centering
        \includegraphics[width=1\linewidth]{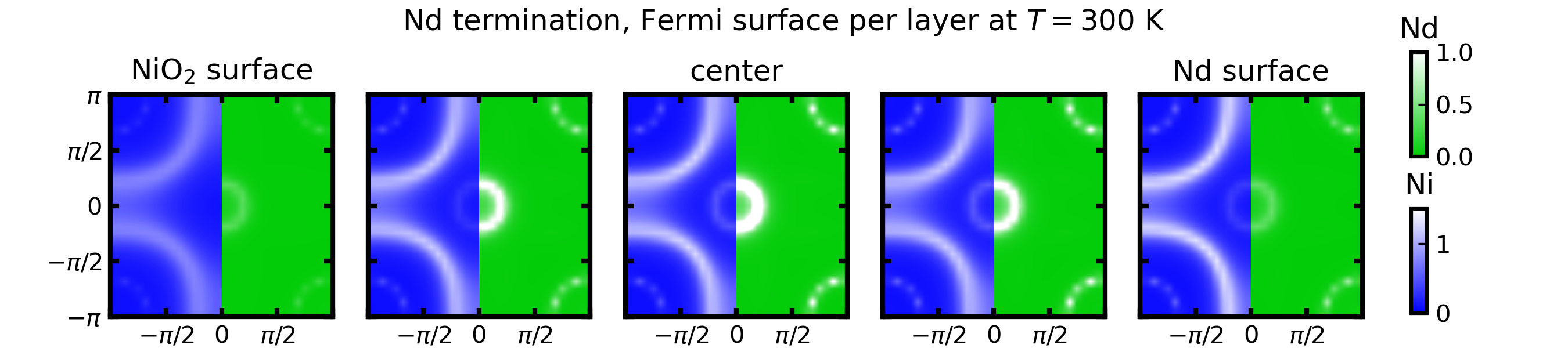}
        \caption{\label{fig:tb_FS_V0}Same as \ref{fig:FS_layer_7_NiO2} for TB model without an external potential.}
    \end{figure}

\begin{figure} [tb]
        \centering
        \includegraphics[width=1\linewidth]{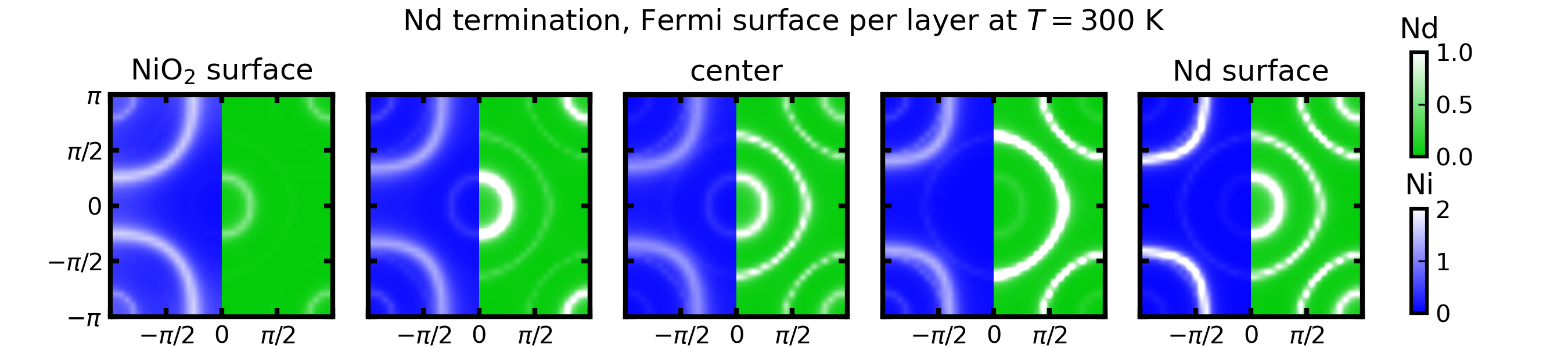}
        \caption{\label{fig:tb_FS_V0.77} Same as \ref{fig:tb_FS_V0} for TB model with an external potential of $V=0.77$\,eV.}
    \end{figure}

We perform DMFT calculations with the same parameters as described in the main text. The Nd and Ni spectral function in the two surface regions differ from those in the bulk region due to their truncated hopping in $z$-direction. However, in Fig.~\ref{fig:tb_dmft} we do not observe the emergence of an insulating layer. This shows: cutting off the hopping to the vacuum alone is
not sufficient to turn the surface layer insulating. Consequently, all layers contribute to the Fermi surface without and with polar field in Fig.~\ref{fig:tb_FS_V0} and \ref{fig:tb_FS_V0.77}, respectively. 

The main aspect missing in this tight-binding modeling, compared to the calculation without substrate in Section~\ref{Sec:NoSTO}, is the tilting of the surface and interface layer out of an ideal planar NiO$_2$ configuration.  This tilting will change the crystal field and is thus obviously important for the depopulation of the Ni $d_{xz/yz}$ orbitals in Fig.~\ref{fig:noSTO} above.
We do not have this multi-orbital physics in  Fig.~\ref{fig:tb_dmft} and thus
the NiO$_2$-surface layer (leftmost in Fig.~\ref{fig:tb_FS_V0.77}) is metallic, whereas it is insulating in Fig.~\ref{fig:FS_layer_noSTO}.


\end{document}